\documentclass{aastex}

\begin{document}

\title{Violation of Richardson's Criterion via Introduction of a \\ Magnetic Field}

\author{Daniel Lecoanet\altaffilmark{1}}
\affil{Department of Physics, University of Wisconsin, Madison, WI 53706 USA}
\email{lecoanet@wisc.edu}
\author{Ellen G. Zweibel\altaffilmark{1}}
\affil{Departments of Astronomy and Physics, University of Wisconsin, Madison, WI 53706 USA}
\author{Richard H. D. Townsend\altaffilmark{1}}
\affil{Department of Astronomy, University of Wisconsin, Madison, WI 53706 USA}
\and
\author{Yi-Min Huang\altaffilmark{1,2}}
\affil{Space Science Center, University of New Hampshire, Durham, NH 03824, USA}

\altaffiltext{1}{Center for Magnetic Self-Organization in Laboratory and Astrophysical Plasmas}
\altaffiltext{2}{Center for Integrated Computation and Analysis of Reconnection and Turbulence}

\begin{abstract}
Shear flow instabilities can profoundly affect the diffusion of
momentum in jets, stars, and disks. The Richardson criterion gives a sufficient
condition for instability of a shear flow in a stratified medium.  The
velocity gradient $V'$ can only destabilize a stably stratified medium
with squared Brunt-V\"{a}is\"{a}l\"{a} frequency $N^2$ if
$V'^2/4>N^2$.  We find this is no longer true when the medium is a
magnetized plasma.  We investigate the effect of stable stratification
on magnetic field and velocity profiles unstable to magneto-shear
instabilities, i.e., instabilities which require the presence of both
magnetic field and shear flow.  We show that a family of profiles
originally studied by \citet{td06} remain unstable even when
$V'^2/4<N^2$, violating the Richardson criterion.  However, not all
magnetic fields can result in a violation of the Richardson criterion.
We consider a class of flows originally considered by \citet{kent68},
which are destabilized by a constant magnetic field, and show that
they become stable when $V'^2/4<N^2$, as predicted by the Richardson
criterion.  This suggests that magnetic free energy is required to
violate the Richardson criterion.  This work implies that the
Richardson criterion cannot be used when evaluating the ideal
stability of a sheared, stably stratified, and magnetized plasma. We
briefly discuss the implications for astrophysical systems.
\end{abstract}

\keywords{instabilities; MHD; stars: magnetic fields; stars: rotation; Sun: magnetic fields; Sun: rotation}

\section{Introduction}\label{sec:intro}

Rotation plays an important role in the structure and evolution of stars.
Although rotation directly modifies hydrostatic equilibrium only in the most rapid
rotators, it drives large scale circulation, modifies the structure of convection and
the nature of convective transport, and is a key component of magnetic dynamos. These
phenomena in turn modify the rotation through a complex interplay of nonlinear
processes. 

Shear flow instability is one of the mechanisms through which rotation influences and is influenced by its environment.
The motion associated with the instability generates stresses, which react back on the flow and drive
it toward a stable state. If the amplitude of the unstable perturbations is sufficiently large, the
motions become turbulent. Shear flow instability and shear flow turbulence can amplify magnetic fields
and mix chemical species, in addition to modifying the rotation profile itself.

In the case of the Sun, and possibly other low mass main sequence stars, the most likely venue for shear
flow instability is the so-called tachocline, the region of strong shear just below the base of the convection
zone \citep[see][for a review]{Gou2007}. Although the mechanisms which maintain the tachocline are still
uncertain, it is almost certainly a component of the solar dynamo, and its existence has
implications for the way the convection zone, which is spun down by the solar wind, is coupled to the
radiative core. The tachocline may be subject to purely hydrodynamic instabilities (\citet{Ras2008}, \citet{KitRud2009}),
global MHD instabilities driven by the latitudinal structure of the field (\citet{GilFox1997}, \citet{Gil2007})
magnetorotational instabilities \citep{Ogi2007}, and, if hydromagnetic forces are
large enough, magnetic buoyancy instabilities
(\citet{Sil2009}, \citet{VasBru2009}).  All these instabilities could modify the tachocline's 
structure.

Massive stars, which evolve quickly and tend to rotate rapidly, are potentially more profoundly
affected by shear flow instability.
The past two decades have witnessed significant advances in
understanding how the internal rotation of massive luminous stars
shapes, and is shaped by, their evolution \citep[see][and references
therein for a comprehensive review]{MaeMey2000}. Rapidly rotating
massive stars follow bluer, more-luminous evolutionary tracks in the
Hertzsprung-Russell diagram (HRD) than non-rotating equivalents, 
because strong meridional circulation injects
fresh hydrogen fuel into the convective core \citep[see,
e.g.,][]{MeyMae2000}. This rotational mixing brings CNO-cycle
nucleosynthetic products from the stars' cores to their surfaces,
leading to changes in photospheric abundance
ratios \citep[e.g.,][]{Tal1997}. 

The prevailing view of rotation in massive stars is based on
a canonical narrative developed by
\citet{Zah1992}. In this scenario, turbulent diffusion of angular momentum is highly anisotropic,
with much stronger transport in the horizontal direction than the
radial one. This leads to a `shellular' rotation profile, in which the
angular velocity is constant on spherical shells. The exchange of
angular momentum between these shells is then mediated by a
combination of meridional circulation, convection (in convective
zones) and radial turbulent diffusion. The turbulence itself is driven
by secular shear instability \citep{MaeMey2000}, which grows
on a thermal timescale \citep[see
also][]{Mae1995,MaeMey1996,TalZah1997}.

Recent studies have considered the role that magnetic fields might
play in modifying angular momentum
transport \citep[e.g.,][]{MaeMey2004}. 
Generally, these studies of the impact of magnetic fields have focused
around contributions to the radial angular momentum diffusivity
arising from the field stiffness \citep{Pet2005}. However,
as \citet{Spr2000} has discussed, a field can also introduce new
instabilities that play a role in angular momentum transport. In this
paper we explore a hitherto-overlooked magnetic-mediated instability,
whereby the presence of a \emph{horizontal} field can destabilize a
stratified shear layer that --- according to the Richardson criterion
--- would otherwise be stable.

The paper is organized as follows.  First we will briefly discuss shear flow instabilities in \S 
\ref{sec:shearintro}.  In \S \ref{sec:basiceqn}, we set up the eigenvalue problem which determines the linear 
stability of an MHD shear flow in a stratified medium.  
We review previous analytic results in \S \ref{sec:analytic}, and describe our numerical methods for solving the eigenvalue problem in \S \ref{sec:numerical}.  Starting in \S \ref{sec:TD} we examine specific examples, first adding stratification to the linear velocity and parabolic magnetic field example considered 
in a recent paper by Tatsuno \& Dorland (2006) (hereafter TD06).  
Our key result is that sufficiently strong parabolic magnetic fields can yield instability for arbitrarily strong stratification, in violation 
of the Richardson criterion.  We consider and extend a family of
 velocity profiles which Kent (1968) (hereafter K68) showed can be destabilized by a constant magnetic field in \S \ref{sec:kent}.  
Contrary to the parabolic magnetic field case, it seems that the introduction of a constant magnetic field cannot 
result in a violation of the Richardson criterion.  This suggests that the free energy of an inhomogeneous magnetic 
field is essential to breaking the Richardson criterion.  We discuss
possible applications to rotating stars in \S \ref{sec:stellar} and conclude in \S \ref{sec:conclusion}.

\section{Introduction to Shear Flow Instabilities}\label{sec:shearintro}

The best known shear flow instability is the hydrodynamic
Kelvin-Helmholtz instability.  The Kelvin-Helmholtz instability has
been studied extensively.  Perhaps the most famous result is the
inflexion point criterion, stating that a necessary condition for
instability is the presence of an inflexion point in the velocity
profile (see, for example, \citet{dr81}).  Others have also given
necessary conditions for instability, making extra assumptions on the
flow profile \citep{lin55, howard61, rs64}.

Many have worked to extend parts of these results to
magnetohydrodynamic shear instabilities.  It is well known that a
sufficiently strong magnetic field stabilizes the Kelvin-Helmholtz
instability \citep{chandra61}.  It was shown years ago, but is perhaps
less well known, that a magnetic field can destabilize an otherwise
stable shear flow (K68). In particular, an inflexion point is no longer
necessary for shear instability.  In hydrodynamics vorticity is frozen into the flow, ensuring that perturbations are stable when there is no
inflexion point \citep{lin55}, but the presence of a magnetic field
can break the vorticity frozen-in condition, relaxing the inflexion
point criterion. K68 constructed a family of flow profiles which are
marginally stable in the absence of a magnetic field and destabilized
by a uniform field parallel to the direction of flow.  TD06 studied
how a linear flow profile, which has no inflexion point and is
marginally stable, can be destabilized by a particular family of
magnetic field profiles.  In particular, TD06 find that a parabolic
magnetic field can render a linear velocity profile unstable.

In this paper, we add a new piece of physics to the analysis: density
stratification.  We employ the Boussinesq approximation and assume
that the plasma is stably stratified, i.e. the squared
Brunt-V\"{a}is\"{a}l\"{a} frequency, $N^2$, is positive.  In
hydrodynamics, the Richardson criterion provides a sufficient
condition for the stability of a shear flow in a stratified medium
(see, for example, \citet{dr81}).  The interchange of two fluid
elements at different heights can release kinetic energy from the
flow.  A necessary condition for instability is that the gravitational
energy required for the interchange must be less than the kinetic
energy released. However, in the presence of an inhomogeneous magnetic
field, energy can also be extracted from the magnetic field, even if
the field would be stable in the absence of shear flow.  Our main
result is that the Richardson criterion no longer holds for
inhomogeneous magnetic fields.

We will only consider the effect of stable stratification on magneto-shear instabilities.  However, \citet{TYM03} studied how a shear flow can destabilize a homogeneous magnetic field in the presence of an \textit{unstable} density gradient.  They found that a linear (Couette) velocity profile can be destabilizing when the velocity shear was not too strong.  Their result is similar to our own in the sense than the system is maximally destabilized when the velocity gradient, magnetic field, and density stratification all have comparable strength.

In this paper we consider only ideal instabilities, i.e. we set the
resistive, viscous, and thermal diffusivities to zero. Diffusive
effects could unleash a host of additional instabilities such as
tearing modes \citep[e.g.,][]{FKR63}, doubly diffusive modes \citep[e.g.,][]{SR83}, and secular shear instabilities \citep[e.g.,][]{MM00}. Although such
instabilities are important in their own right, in this paper we
focus entirely on dynamical instabilities.

\section{Basic Equations}\label{sec:basiceqn}

The time evolution of an ideal, incompressible plasma is given by
\begin{eqnarray}\label{eq:fullequations1}
\rho\left(\frac{\partial \mathbf{V}}{\partial t}+\mathbf{V}\cdot\mbox{\boldmath $\nabla$}\mathbf{V}\right) &=& -\mbox{\boldmath $\nabla$}\left(p+\frac{B^2}{2\mu_0}\right)+\frac{1}{\mu_0}\mathbf{B}\cdot\mbox{\boldmath $\nabla$}\mathbf{B}-g\rho\mathbf{e}_z, \\
\label{eq:fullequations2}\frac{\partial\mathbf{B}}{\partial t} &=& \mbox{\boldmath $\nabla$}\times\left(\mathbf{V}\times\mathbf{B}\right), \\
\label{eq:fullequations3}0 &=& \mbox{\boldmath $\nabla$}\cdot \mathbf{V}, \\
\label{eq:fullequations4}0 &=& \mbox{\boldmath $\nabla$}\cdot\mathbf{B}, \\
\label{eq:fullequations5}0 &=& \frac{\partial\rho}{\partial t}+\mathbf{V}\cdot\mbox{\boldmath $\nabla$}\rho,
\end{eqnarray}
where the symbols have their usual meanings.  Equation
(\ref{eq:fullequations1}) is the momentum equation,
eqn. (\ref{eq:fullequations2}) is the induction equation,
eqn. (\ref{eq:fullequations3}) enforces incompressibility,
eqn. (\ref{eq:fullequations4}) is the divergenceless magnetic field
condition, and eqn. (\ref{eq:fullequations5}) is the continuity
equation.  We will write the unit vectors in the $x, y,$ and $z$
directions as $\mathbf{e}_x$, $\mathbf{e}_y$, and $\mathbf{e}_z$
respectively.  The gravitational strength is parameterized by $g$, and
gravity is assumed to point in the $- \mathbf{e}_z$ direction.  We denote
background velocity and magnetic fields with capital letters, and then
perturb the background fields with fields denoted with lower case
letters, except that the background density is denoted $\rho$, and the
perturbed density $\hat{\rho}$.  We assume that the background
quantities $\rho, \mathbf{V},\mathbf{B}$ are all functions of only
$z$, and that our domain is the volume between $z=-z_0$ and $z=+z_0$
with ``free-slip,'' perfectly-conducting boundary conditions in the
$z$ direction, and periodic boundary conditions in the $x$ and $y$
directions.  By ``free-slip,'' we mean no constraint on perturbed
quantities in the $x$ and $y$ directions at the walls, but that
perturbations have no $z$ component at the walls. These are the
boundary conditions adopted by TD06 (who termed them ``no-slip'' which
is not correct --- as will be shown in \S
\ref{sec:TDeigenfunction}, the perturbations slip along, but do not
penetrate, the walls).  Next, we assume that $\mathbf{V}$ is oriented in
only one direction throughout the domain, which we define to be the
$x$ direction.  Thus, we take
\begin{equation}\label{eq:velocity}
\mathbf{V}=(V(z),0,0),
\end{equation}
in Cartesian coordinates.  The background magnetic field $\mathbf{B}$ is
\begin{equation}\label{eq:magneticfield}
\mathbf{B}=(B_x(z),B_y(z),0),
\end{equation}
in Cartesian coordinates.  The background fields are assumed to be in equilibrium, so we have that
\begin{equation}\label{eq:pressure}
\mbox{\boldmath $\nabla$}\left(p+\frac{B^2}{2\mu_0}\right)+g\rho\mathbf{e}_z =0.
\end{equation}
Equation (\ref{eq:pressure}) specifies an integral equation for the
background pressure $p$ for arbitrary $B$ and $\rho$.  The induction
equation and continuity equation for the background fields are
automatically satisfied by the geometry we have imposed.

Now assume the perturbation fields all have the form
\begin{equation}\label{eq:fouriertransform}
f(x,y,z,t)=f(z)\exp(ik_x x+ik_y y-ik_xct).
\end{equation}
We will take $\mathbf{k}\equiv k_x\mathbf{e}_x+k_y\mathbf{e}_y$, $k=|\mathbf{k}|$ and
$\mathbf{\hat{k}}=\mathbf{k}/k$.  In many applications, the density
gradient $\rho'$ is small in comparison to the velocity gradient $V'$
--- where prime denotes differentiation with respect to $z$ --- but
the strength of gravity $g$ is large.  Assuming this, we recover the
Boussinesq approximation, in which we drop terms proportional to
$\rho'$ alone, but keep terms proportional to $g\rho'$.  These
assumptions yield the following eigenvalue problem for $\xi$, the
plasma displacement in the $z$ direction:
\begin{equation}\label{eq:eigeqnkyfinite}
\left(\left[k_x^2\left(V-c\right)^2 - k^2 A^2\right]\xi'\right)' - k^2\left[k_x^2 \left(V-c\right)^2 - k^2 A^2\right]\xi +k^2 N^2 \xi = 0,
\end{equation}
where $A\equiv \mathbf{\hat{k}}\cdot\mathbf{B}/\sqrt{\rho\mu_0}$ is
the Alfv\'{e}n velocity, and $N^2\equiv g\rho'/\rho$ is the
Brunt-V\"{a}is\"{a}l\"{a} frequency in the Boussinesq
approximation. To simplify our analysis, we assume that $N^2$ is constant throughout
the domain, which corresponds to the exponentially decaying density
profile. When computing the Alfv\'{e}n velocity, the Boussinesq approximation will allow us to consider $\rho$ to be a constant. The boundary conditions are that
$\xi=0$ at the boundaries at $z=-z_0$ and $z=+z_0$.

There is an asymmetry in how velocity shear, magnetic fields, and density stratification depend on the wavenumber $\mathbf{k}$.  For $\mathbf{k}=k_y\mathbf{e}_y$, $k_x=0$ and the velocity shear is irrelevant (note that $k_xc$, the growth rate, could still be finite).  The purpose of this paper is to examine the interplay between velocity and magnetic fields, so we will not consider this case.  Also note that the Alfv\'{e}n velocity, as it occurs in eqn. (\ref{eq:eigeqnkyfinite}), is a function of $\mathbf{\hat{k}}$.  For example, if $\mathbf{B}$ is constant in the $z$ direction, there exists a $\mathbf{\hat{k}}$ for which $A=0$, so the magnetic field would have no effect on such a perturbation.  The strength of gravity in relation to shear flow contains a factor of $k^2/k_x^2$.  Thus, gravity is maximally destabilized by shear flows when $k_y=0$.

Consider an eigenvalue problem for magnetic field $\mathbf{B}$, velocity $V$, Brunt-V\"{a}is\"{a}l\"{a} frequency $N^2$, and wavenumber $\mathbf{k}=k_x\mathbf{e}_x+k_y\mathbf{e}_y$, with $k_y\neq 0$.  We will show that this eigenvalue problem is equivalent to another eigenvalue problem with $k_y=0$, but with different $\mathbf{B}$, $N^2$, and $k_x$.  Define $\mathbf{B}'\equiv\mathbf{e}_x\mathbf{k}\cdot\mathbf{B}/k_x$, $N'^2\equiv k^2N^2/k_x^2$, and $\mathbf{k}'\equiv k \mathbf{e}_x$.  Then the magnetic field $\mathbf{B}'$, velocity $V$, Brunt-V\"{a}is\"{a}l\"{a} frequency $N'^2$, and wavenumber $\mathbf{k}'$ have the same eigenvalue equation as above.  Thus, finite $k_y$ is equivalent to $k_y=0$, if one appropriately rotates and augments the magnetic field, and increases the density stratification.  With this in mind, we will consider the $k_y=0$ case in the remainder of this paper, for which the eigenvalue equation reduces to

\begin{equation}\label{eq:eigeqn}
\left(\left[\left(V-c\right)^2 - A^2\right]\xi'\right)' - k^2\left[\left(V-c\right)^2 - A^2\right]\xi +N^2 \xi = 0.
\end{equation}

The eigenvalue eqn. (\ref{eq:eigeqn}) possesses some symmetries.
First, the sign of $A$ is unimportant, so changing the sign of the magnetic field
does not change the problem.  Another symmetry is translational:
taking $V\rightarrow V+\Delta V$ and $c\rightarrow c-\Delta V$
corresponds to Galilean transformations.  Thus, without loss of
generality, we can and do put ourselves in a frame in which $V(0)=0$.
To make the problem more tractable, we add additional symmetries to
the equation by postulating that $A$ is even in $z$ and $V$ is odd. There is a rescaling symmetry: eqn. (\ref{eq:eigeqn}) remains invariant under
\begin{eqnarray}\label{eq:rescale}
z\rightarrow z/z_0,\nonumber \\
V\rightarrow V/z_0,\nonumber \\
A\rightarrow A/z_0, \\
k\rightarrow kz_0, \nonumber \\
c\rightarrow c/z_0. \nonumber
\end{eqnarray}
Note that $N^2$ and $\mbox{Ri}$ are left unchanged under this transformation.

There is also structure in the eigenvalues.  In general, $c$ and $\xi$
are complex; $c=c_r+ic_i$, $\xi=\xi_r+i\xi_i$.  We ignore the singular
$c_i=0$ case.  If $\xi_c$ is an eigenfunction with eigenvalue $c$,
then $\xi_c^*$, the complex conjugate of $\xi_c$, is a solution to
eqn. (\ref{eq:eigeqn}) with eigenvalue $c^*$.  Thus, eigenvalues come
in complex conjugate pairs, regardless of the symmetry properties of
$A$ and $V$. Assuming that $A$ is even and $V$ is odd, we can show
that if $c$ is an eigenvalue, then $-c$ is also an eigenvalue, with
eigenfunction $\xi_c(-z)$.

Numerically, we only find eigenvalues with $c_r=0$, and with the
following eigenfunction symmetry: If we normalize the eigenfunction
$\xi$ such that $\xi(0)=1$, then $\xi_r$ is even and $\xi_i$ is odd.
In \S\S \ref{sec:TD} and \ref{sec:kent}, we assume that $c_r=0$
and the eigenfunction has this symmetry.  These properties are
linked. If we multiply eqn. (\ref{eq:eigeqn}) by $\xi^*$ and integrate
over the domain the result is
\begin{equation}\label{eq:integrated}
\int_{-z_0}^{z_0}\left[(V-c)^2-A^2\right]\left(\left|\frac{d\xi}{dz}\right|^2+k^2\left|\xi\right|^2\right)-N^2\vert\xi\vert^2dz=0.
\end{equation}
The imaginary part of eqn. (\ref{eq:integrated}) is
\begin{equation}\label{eq:ip}
2ic_i\int_{-z_0}^{z_0}(c_r-V)\left(\left|\frac{d\xi}{dz}\right|^2+k^2\left|\xi\right|^2\right)dz=0.
\end{equation}
If the real and imaginary parts of $\xi$ each have definite parity,
the term proportional to $V$ in eqn.  (\ref{eq:ip})
vanishes. Therefore $c_rc_i\equiv 0$, and unstable modes have
$c_r=0$. This result is useful in searching for unstable modes, as
described in \S\ref{sec:numerical}.

We find that generally the growth rate $c=ic_i$ is small in comparison
to $V$, which is ${\mathcal{O}}(1)$.  When $V^2=A^2$, the coefficient of
the $\xi''$ term in eqn. (\ref{eq:eigeqn}) goes to $|-c_i^2|\ll 1$.
Thus, the equation becomes ``almost singular'' when $|V|=|A|$, and
becomes actually singular when $c=0$.  The ``almost singularities''
are characterized by large gradients in the eigenfunctions, as is
shown in \S\S\ref{sec:TD} and \ref{sec:kent}.

We will often consider the limit $k^2=0$.  When $k^2=0$, the growth rate,
$kc$, is formally zero.  However, one can view the eigenvalue $c$ as a
function of the various parameters $A, V, k^2, N^2$.  We assume that
$c(k^2)$ is analytic about $k^2=0$, so our results for the $k^2=0$
case still hold in a neighborhood of $k^2=0$.  Thus, when we consider
$k^2=0$, we are really taking the limit as $k$ becomes small.  The
$k^2$ term in eqn. (\ref{eq:eigeqn}) is only important when it is
comparable to the scale heights of the velocity and magnetic fields
and the perturbation $\xi$.  Numerically, we find that $kz_0<0.1$ is
``small'' for the examples presented in this paper.

\section{Review of Analytic Results}\label{sec:analytic}

Shear flow instabilities are global instabilities.  Thus, the two
categories of analytic results --- necessary conditions for
instability and sufficient conditions for instability --- can be
viewed as local and global conditions.  Necessary conditions for
instability give criteria which must be satisfied in at least one spot
in the domain, whereas the sufficient conditions for instability are
global criteria involving integrals over the domain.  We present a
short overview of the analytic results regarding the linear stability
of shear flows.  We begin by discussing shear flows alone, and then
add stratification, a magnetic field, and then both.  The zero
magnetic field and zero density gradient cases can be viewed as limits
of the more general problem.

\subsection{Shear Flow Instabilities}\label{sec:shear}

Probably the best known result is the inflexion point criterion, which
states that $V''$ must have a zero in the domain for there to be
instability.  This is a local, necessary condition.  There are several
physical interpretations of the inflexion point criterion.  Consider
the Reynolds stress of the perturbation,
$\tau=-\rho\overline{v_xv_z}$, where the bar denotes averaging with
respect to $x$.  Assuming $c\neq 0$, one can show that $d\tau/dz$ has
a zero iff $V''$ has a zero (for instance, in \citet{lin55} or K68).
Since $\tau=0$ at the boundaries, when $c\neq 0$, we must have that
$V''$ has a zero. \citet{lin55} has proposed an alternate
interpretation considering vorticity.  A zero in $V''$ corresponds to
an extremum in vorticity, and Lin has shown that perturbations feel a
restoring force unless they are at an extremum of vorticity.

The inflexion point theorem is useful because it rules out a large
class of velocity profiles as stable.  However, it cannot be used to
show that a particular shear flow is unstable.  \citet{rs64} were able
to prove a necessary and sufficient condition for instability by using
the additional assumptions that $V''$ has a single zero and $V$ is
monotonic.  Under these assumptions, $V$ is unstable in $z_1\le z\le
z_2$ if and only if
\begin{equation}\label{eq:RS}
\left.\frac{1}{V'(V_c-V)}\right|^{z_2}_{z_1}-\int_{z_1}^{z_2} \frac{V''}{V'^3(V-V_c)}dz>0,
\end{equation}
where $V_c$ is the velocity at the inflexion point.  This result is
derived for the $k^2=0$ case.  A priori, it seems that there could be
velocity profiles which are unstable for $k^2>0$ but stable for
$k^2=0$.  Then an instability condition for $k^2=0$ would be only
sufficient for instability.  This is addressed by a theorem of
\citet{lin55} which shows that under the assumptions of Rosenbluth \&
Simon, velocity profiles which are unstable for $k^2>0$ are also
unstable for $k^2=0$.

\subsection{Shear Flow Instabilities in a Stratified Medium}\label{sec:shearstrat}

The key stability result for stratified media is the Richardson
criterion, a necessary condition for the instability of a shear flow
in a stratified medium.  If
\begin{equation}\label{eq:RC}
\mbox{Ri}\equiv\frac{N^2}{V'^2} >\frac{1}{4}
\end{equation}
everywhere, then there is stability.  A physical interpretation (see,
for example, \citet{chandra61} or \citet{dr81}) is that if exchanging
fluid elements at slightly different heights increases the potential
energy more than it decreases the kinetic energy, then the
perturbation is stable.

Provided that $\mbox{Ri} < 1/4$, we have that
\begin{equation}\label{eq:cbound}
k^2c_i^2\leq \max \left(\frac{1}{4}V'^2-N^2\right).
\end{equation}
This result by \citet{howard61} follows from the proof of the
Richardson criterion and is also discussed in \citet{dr81}).

\subsection{Magneto-Shear Instabilities}\label{sec:magnetoshear}

Magnetic fields can both stabilize and destabilize shear flows.  First
we consider their stabilizing effect.  Perturbations which bend
magnetic fieldlines induce a restoring magnetic tension force.  A
classic result is that in a constant density medium, the vortex sheet
$V(z)=-U$ for $z<0$ and $V(z)=+U$ for $z>0$ for some constant $U$, is
stabilized by a magnetic field $A$ if and only if $A^2>V^2$
\citep{chandra61}.  This step function velocity profile is the
limiting distribution of $V(z)=U_0 \tanh (z/a)$ as $a\rightarrow 0$.
\citet{keppens99} have investigated the hyperbolic tangent $V$ case
with a constant magnetic field, including compressibility, and found
the magnetic field stabilizing.  These results were qualitatively
similar to those by Chandrasekhar, which is expected because a
constant magnetic field has no length scale (or it has an infinite
length scale), so it cannot tell the difference between the
$a\rightarrow 0$ and $a$ finite case.

Keppens et al also found that the addition of a non-uniform magnetic
field could be destabilizing.  When they added a small field
$A(z)=-A_0$ for $z<0$ and $A(z)=A_0$ for $z>0$, they found that the
growth rate increased, and was even larger when $A$ reversed
smoothly. Although their calculation, unlike ours, includes
compressibility, there is one robust effect which is always present:
magnetic fields allow transfer of vorticity between fluid
elements. The loss of the frozen-in vorticity constraint changes the range of
motions allowed in the plasma, and yielding instability.

We now review some general results on magnetoshear instabilities in
order to understand how the Richardson criterion can be violated by
the introduction of a magnetic field.

The necessary and sufficient instability condition of \citet{rs64}
(eqn. (\ref{eq:RS})) has been generalized to the MHD case by K68 and
\citet{cm91}. Both arguments use that when $k^2=0$, there is an exact
solution to eqn. (\ref{eq:eigeqn}),
\begin{equation}\label{eq:exactsolution}
\xi (z)=\int_{z_1}^{z} \frac{dz'}{(V-c)^2-A^2},
\end{equation}
and then define
\begin{equation}\label{eq:definef}
f(c)\equiv\int_{z_1}^{z_2} \frac{dz}{(V-c)^2-A^2}=\xi(z_2).
\end{equation}
The eigenvalues of eqn. (\ref{eq:eigeqn}) are then just the zeros of
$f(c)$, and one can search for instabilities by implementing Nyquist's
method to determine if there are any zeros of $f(c)$ for $c_i>0$.
Nyquist's method is an application of the Argument Principle (see, for
instance, \citet{gamelin01}), which states that the integral of the
argument of $f(c)$ on the boundary $\partial D$ of some region $D$ is
equal to $2\pi(N_0-N_\infty)$, where $N_0$ is the number of zeros of
$f(c)$ in $D$ and $N_\infty$ is the number of poles of $f(c)$ in $D$.
In our case, we assume $N_\infty=0$, so counting the number of times
$f(c)$ wraps around the origin tells us how many zeros, i.e. unstable
modes, there are.  Further discussion of Nyquist's method can be found
in \citet{kt73}.

Nyquist's method can only be applied if we know what contour
to use.  The real part of $c$ can be bounded by extending an important
hydrodynamic result by Rayleigh.  It can be shown \citep{ht01} that
$c_r$ must lie in the range of $V$, so the contour in $c$ space is
bounded by $V_{min}<c_r<V_{max}$.  The lower bound for $c_i$ is $0^+$,
and the upper bound can be recovered by modifying Howard's semicircle
theorem \citep{howard61}.  In the hydrodynamic case, Howard showed
(see, for instance, \citet{dr81}) that
\begin{equation}\label{eq:semicircle}
\left[c_r-\frac{1}{2}\left(V_{max}+V_{min}\right)\right]^2+c_i^2\leq \left[\frac{1}{2}\left(V_{max}-V_{min}\right)\right]^2.
\end{equation}
Thus, we have that $c_i\leq 1/2 (V_{max}-V_{min})$.  \citet{ht01} have
shown that in MHD, we have the two inequalities
\begin{equation}\label{eq:msemicircle1}
(V^2-A^2)_{min}\leq c_r^2+c_i^2\leq (V^2-A^2)_{max},
\end{equation}
and
\begin{equation}\label{eq:msemicircle2}
\left[c_r-\frac{1}{2}\left(V_{max}+V_{min}\right)\right]^2+c_i^2\leq \left[\frac{1}{2}\left(V_{max}-V_{min}\right)\right]^2-\left(A^2\right)_{min}.
\end{equation}
This gives an even stronger upper bound on $c_i$, that
\begin{equation}\label{eq:cbound2}
c_i\leq\sqrt{(1/2(V_{max}-V_{min}))^2-(A^2)_{min}}.
\end{equation}
These two inequalities can be used to show stability, if one can show
that there are no $c$ which simultaneously satisfy both inequalities.

\citet{cm91} used Nyquist's method to provide a sufficient condition
for instability for flows in which $V$ is even, and $A$ is either odd
or even. They showed that
\begin{equation}\label{eq:CMcondition}
\Re \int_{-z_0}^{z_0}\frac{dz}{(V-i\epsilon )^2-A^2}>0
\end{equation}
as $\epsilon \rightarrow 0$ is sufficient for instability.  Note that
it is not assumed that $V$ has an inflexion point.

K68 considered the effects of a small, constant magnetic field on a
stable velocity profile.  He showed that when $V''$ has a single zero,
and there exist points $y_s, y_t$ such that the velocities at these
points, $V_s, V_t$ satisfy $V_s-V_t=2A$ and $V'_s-V'_t=0$, then
\begin{equation}\label{eq:kentcondition}
M(A)\equiv \wp\int_{z_1}^{z_2} \frac{dz}{(V-c_0)^2-A^2}>0,
\end{equation}
implies instability.  Here, $c_0$ is defined by $c_0=(V_s+V_t)/2$, and $\wp$ denotes the
principal value of the integral.  For small $A$,
$c_0$ is the velocity at the inflexion point, but as $A$ increases, it
can deviate somewhat.  For a marginally stable velocity profile, we
have $M(0)=0$.  In the remainder of this section, we will use $\dot{ }$ (dot) to denote derivative with respect to $A$.  In the limit $A\rightarrow 0$, we have
$\dot{M}(A)\rightarrow 0$.  Thus, to evaluate the stability of $V$ to
infinitely small $A$, we need to consider $\ddot{M}(0)$, which Kent
shows is given by
\begin{equation}\label{eq:mdoubledot}
\ddot{M}(0)=2\ddot{c}_r(0)\int_{z_1}^{z_2}\frac{dz}{(V-V_0)^3}+2\int_{z_1}^{z_2}\frac{dz}{(V-V_0)^4},
\end{equation}
where $V_0$ is the velocity at the inflexion point and
\begin{equation}\label{eq:crdoubledot}
\ddot{c}_r(0)=-\frac{V_0^{(4)}}{3V_0V_0^{(3)}}.
\end{equation}
This criterion is useful because one can change variables to integrate
over $V$, and if $V_0=0$ and $\omega (V):=dz/dV$ is even, then
\begin{equation}\label{eq:kentcondition2}
\ddot{M}(0)=\int_{V_1}^{V_2} \frac{\omega dV}{V^4},
\end{equation}
where $V_i=V(z_i)$.  Although these conditions are sufficient for
instability, they are not necessary.  Unlike in the hydrodynamic case,
there can be unstable modes for finite $k^2$ for a velocity profile
which is stable at $k^2=0$ (K68).

Another way to tackle the general problem with arbitrary velocity and
magnetic field profiles is to attempt to extend the physical arguments
behind the inflexion point criterion to the MHD problem.  In the MHD
problem, one must consider both the Reynolds and Maxwell stresses, so
the total stress is given by
\begin{equation}\label{eq:totalstress}
\tau_{tot}=-\rho\overline{v_xv_z}+\overline{b_x b_z}.
\end{equation}
A necessary condition for instability is still $d\tau_{tot}/dz=0$
somewhere in the flow. K68 has shown that this condition can be
written as
\begin{equation}\label{eq:kentinf1}
\Im\left[ |X|\frac{X'}{X} \right]'=0,
\end{equation}
or
\begin{equation}\label{eq:kentinf2}
\Im\left[ \frac{2XX''-X'^2}{4X^2} \right]=0,
\end{equation}
where $X\equiv (V-c)^2-A^2$.  Unfortunately, these (equivalent)
conditions are not as useful as the inflexion point criterion because
they depend on both the flow profile and the growth rate.  Thus, one
needs to check eqns. (\ref{eq:kentinf1}) or (\ref{eq:kentinf2}) for
all possible $c$.  This condition seems to be fairly weak, and is
satisfied by many stable profiles.

\subsection{Magneto-Shear Instabilities in a Stratified Medium}\label{sec:ssf}

The addition of a magnetic field to a shear flow in a stratified
medium makes the problem significantly more complex.  The Richardson
criterion is no longer valid, but it can be generalized.  We have
carried out the same analysis used to derive the Richardson criterion,
but included magnetic fields.  The result is that if
\begin{equation}\label{eq:extendedRC}
0>\frac{1}{c_i}\Im\left(\frac{2ZZ''-Z'^2}{4Z^2}+\frac{V'Z'}{Z}+\frac{\frac{V'^2}{4}-\frac{N^2}{Z}}{(V-c)}\right)
\end{equation}
everywhere in the domain, then the system is stable. Here, $Z\equiv
1-A^2/(V-c)^2$. Similarly to the generalization of the inflexion point
criterion (eqn. (\ref{eq:kentinf1}), (\ref{eq:kentinf2})), this
condition involves $c$.  This condition also seems to be weak.

Although we normally assume that $c_r=0$, this condition can be
relaxed, and we can find bounds for $c_r$.  The argument by
\citet{ht01} mentioned in \S \ref{sec:magnetoshear} still holds
when stratification is introduced, and shows that $c_r$ must lie
within the range of $V$.  This bound on $c_r$ is valid with and
without magnetic field, and with and without stratification.

\section{Numerical Methods}\label{sec:numerical}

Because the problem is global, analytic results exist only in cases
with particular symmetries (i.e., $k^2=0$ or $N^2=0$), so we must
generally solve for stability numerically.  We have implemented three
numerical methods for solving the eigenvalue problem,
eqn. (\ref{eq:eigeqn}).  In the first, we discretize the equation onto
a Chebyshev grid, and use a finite dimensional approximation for the
differential operator.  Then eqn. (\ref{eq:eigeqn}) can be rewritten
as a generalized finite-dimensional eigenvalue equation:
\begin{equation}\label{eq:geneigeqn}
\gamma \left(\begin{array}{ccc}
\mathcal{D} & 0 & 0 \\
0 & 1 & 0 \\
0 & 0 & 1 \end{array}\right)
\left( \begin{array}{c}
v_z \\
b_z \\
\hat{\rho} \end{array} \right) =
\left( \begin{array}{ccc}
-ik_xV\mathcal{D}+ik_xV'' & ikA\mathcal{D}-ikA'' & -N^2k^2 \\
ikA & -ik_xV & 0 \\
1 & 0 & -ik_xV \end{array}\right)
\left(\begin{array}{c}
v_z \\
b_z \\
\hat{\rho} \end{array}\right)
\end{equation}
where $\mathcal{D}\equiv \partial_z^2-k^2$.  Matlab was used to solve
this finite dimensional eigenvalue problem.  This approach was useful
when we did not require high resolution.  This method was not able to
resolve the large gradients in the eigenfunctions that sometimes
appeared when $|V|=|A|$.

Another strategy, for $k=0$, was implementing Nyquist's method. We
used Mathematica to calculate $f(c)$, as defined in
eqn. (\ref{eq:definef}) for various $c$.  As mentioned in \S
\ref{sec:ssf}, we know that $c_r$ lies between the minimum and maximum
of $V$.  The advantage of Nyquist's method is that we need not assume
that $c$ is imaginary.  We picked the rectangle with vertices at
$i\epsilon +V_{max}$, $i\epsilon +V_{min}$, $ia+V_{min}$, and
$ia+V_{max}$ as the contour, with $a$ of order one and $\epsilon$
small.  If one plots $f(c)$ where $c$ traverses this contour, it is
easy to see if there are any unstable modes with $c$ in this contour.
We varied the size of the rectangular contour to find the exact
eigenvalues.  For the examples presented below in \S\S
\ref{sec:TD} and \ref{sec:kent}, eigenvalues were always purely
imaginary, and the eigenfunctions had the symmetry properties
described in \S \ref{sec:basiceqn}.

Finally, we used a finite difference relaxation code to integrate
across the domain.  We assumed that $c$ was imaginary, and integrated
eqn. (\ref{eq:eigeqn}) over the domain for $c$ between $i\epsilon$ and
$ia$ for $a$ of order one and $\epsilon$ small, in logarithmic steps.
When the real part of $f(c)$ changed sign between two consecutive
steps, the secant method was used to find the zero in the real part of
$f(c)$, which corresponds to a zero in $f(c)$.  This algorithm was the
most efficient, but makes the assumption that the eigenvalues are
purely imaginary.  As mentioned in \S \ref{sec:basiceqn}, we have
not found any eigenvalues with non-vanishing real part using the other
two methods mentioned above, so this seems to be a valid assumption.

All three numerical methods give similar results in cases where we
used more than one.

\section{Linear $V$, Parabolic $A$}\label{sec:TD}

In this section, we add density stratification to the linear velocity
and parabolic magnetic field profiles considered by TD06.  The main
result is that we find instability even when $V'^2/4<N^2$ everywhere,
i.e., when the Richardson criterion predicts stability.  We believe
this is because the magnetic field provides another free energy source
for the instability.  At $k^2=0$, there are magnetic field profiles
which are unstable for arbitrarily large $N^2$, but when $k^2>0$,
there is only a finite range of $N^2$ which are unstable for the
profiles considered here.

\subsection{The Field and Flow Profiles}\label{sec:TDN2=0}

Consider the following velocity and magnetic field profiles in a
domain from $z=-1$ to $z=+1$:
\begin{eqnarray}\label{eq:TDprofilesV}
V(z)&=&z, \\
\label{eq:TDprofilesA}A(z)&=&(1-\alpha)z^2+\alpha.
\end{eqnarray}
These are the fields considered by TD06 in Section III.A.1 (where we
call their $\alpha_1$ parameter $\alpha$).  The magnetic field is a
parabola with $A(0)=\alpha$ and $A=1$ at the boundaries.

An important characteristic of these profiles is that neither the
magnetic field nor the velocity profile are unstable by themselves.
The instability is truly a magneto-shear instability, as both magnetic
field and shear flow play a part in rendering the profiles unstable.
In this respect, this example is different from those considered by
others in which a magnetic instability is stabilized by gravity
\citep{dg09}, a magnetic layer destabilizes a stratified medium
\citep{n61}, or magnetic field and shear flow modify a buoyancy
instability \citep{h01}.

These profiles can be viewed as local approximations to a wide range
of field and flow profiles. The parabolic magnetic field profile is
valid locally whenever $B$ has an extremum, which we take to be at
$z=0$. As mentioned in \S \ref{sec:basiceqn}, taking
$A\rightarrow -A$ does not change the problem, so although we are
considering a local minimum, the exact same results hold for
$A(z)=-(1-\alpha)z^2-\alpha$, which characterizes a local maximum. We
can always transform to a frame in which $V(0)=0$, so the velocity has
a local expansion of the form of equation (\ref{eq:TDprofilesV}).

To view these profiles as a local approximation, we also need to make
an assumption about the relative strength and scale of
variation of the magnetic field and
the shear flow, since we require that $|V|=|A|$ at the boundary.  When
$\alpha$ is close to zero, the magnetic field and velocity are
changing at similar rates, so the locality assumption is plausible;
but when $\alpha$ is close to one or very negative, the scale heights
of the flow and magnetic field are very different, so viewing these
profiles as a local expansion is not as accurate.

Depending on the sign of $\alpha$, the magnetic field has either two
or zero nulls.  When $\alpha<0$, $A=0$ at
\begin{equation}\label{eq:TDnull}
z=\pm \sqrt{\frac{\alpha}{\alpha-1}}.
\end{equation}
When $\alpha>0$, there are no nulls in the magnetic field, and when
$\alpha=0$, there is a single null at $z=0$.  We find that the nulls
in the magnetic field are unimportant in this problem --- rather,
zeros of $V^2-A^2$ are important.  The eigenfunctions discussed below
(see \S \ref{sec:TDeigenfunction}) show no special behavior at
$A=0$, but have sharp gradients when $|A|=|V|$.  In terms of $\alpha$,
$|V|=|A|$ at
\begin{eqnarray}\label{eq:TDVeqA1}
z&=&\pm 1, \\
\label{eq:TDVeqA2} z&=&\pm \frac{\alpha}{1-\alpha}.
\end{eqnarray}
When $\alpha>0.5$, the solutions in eqn. (\ref{eq:TDVeqA2}) are no
longer in the domain.  This means that $V\leq A$ in the entire domain,
yielding stability by eqn. (\ref{eq:CMcondition}).  Heuristically,
when $\alpha$ becomes more positive, the strength of the magnetic
field in the domain increases until the magnetic tension force becomes
so strong that all perturbations become stable.

In the opposite limit, when $\alpha$ becomes very negative, the
solutions in eqn. (\ref{eq:TDVeqA2}) approach $z=\pm 1$.  For
arbitrarily negative $\alpha$, there is still some region for which
$V>A$.  Tatsuno \& Dorland find instability for $\alpha$ as small as
-25, and we can prove that there is instability for all $\alpha<0.5$
when $k^2=0$ using the sufficient condition for instability by Chen \&
Morrison described in \S \ref{sec:analytic}.  The explicit
computation is messy, but is included in Appendix \ref{sec:alphainf}.

The limit in which $\alpha\rightarrow -\infty$ is probably not
physically relevant.  As the two ``almost singular'' layers  approach
each other (see eqns. \ref{eq:TDVeqA1} and \ref{eq:TDVeqA2}), 
there are large gradients at the boundary of the domain.
In this case, the instability probably relies crucially on our choice
of boundary conditions. Moreover, when stratification is included, the
high field strengths and large currents corresponding to
$\vert\alpha\vert\gg 1$ are destabilizing in themselves, contrary to
what we assume here. Thus, results in this limit should be viewed as
proving a point about the Richardson criterion, but are not
necessarily physically relevant by themselves.  As we show in explicit
calculations presented below, $\alpha$ does not need to be very
negative to recover the results described in the infinitely negative
case.

\subsection{Effect of Stratification on Stability}\label{sec:TDNneq0}

Our main result is evidence for the following conjecture: There is
instability as $\alpha\rightarrow -\infty$, even in the presence of
arbitrarily strong density stratification, in violation of the
Richardson criterion.  There does not seem to be any way to prove this
claim analytically, as there was in the $N^2=0$ case.  The sufficient condition for
stability presented by Chen and Morrison relies crucially on the
analytic solution to the eigenvalue equation when $k^2=0$.  When
$N^2\neq 0$, we no longer have an analytic solution to the eigenvalue
equation, even when $k^2=0$, so there is no extension of the proof.

Given the assumptions made above, the growth rate $c$ is a function of
the following parameters: $k^2$, $N^2$, and $\alpha$.  We first
specialize to the $k^2=0$ case, and then examine the more general
$k^2$ finite case.

\subsubsection{$k^2=0$}\label{sec:TDk=0}

\begin{figure}[!h]
\plotone{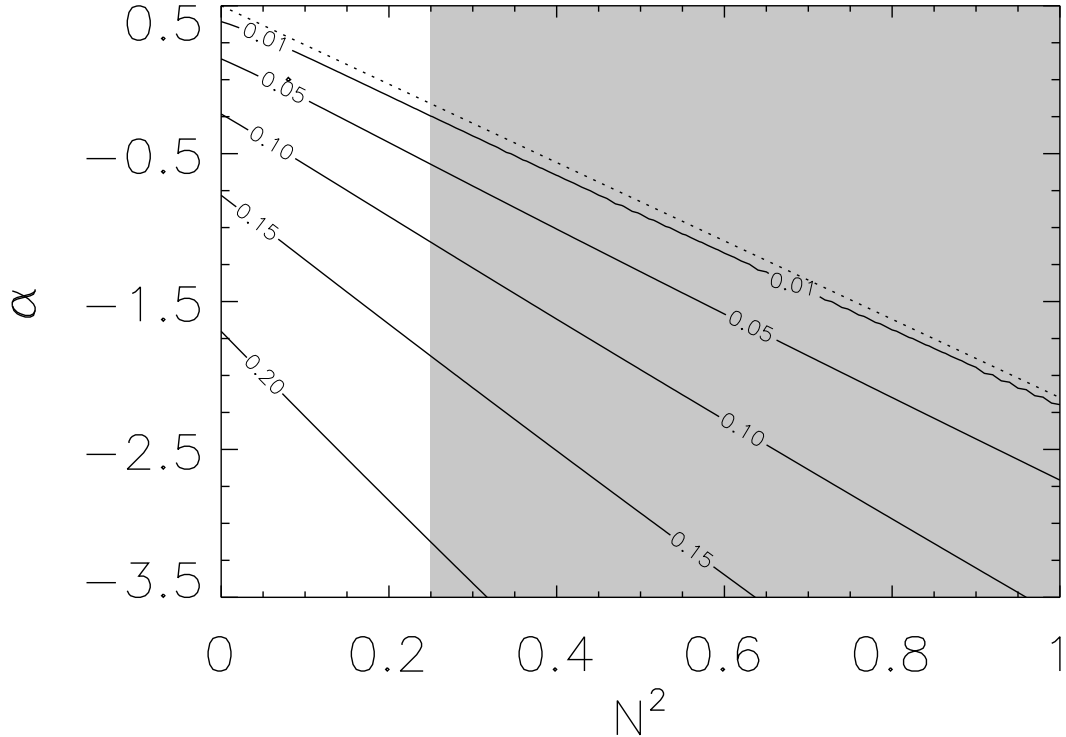}
\caption{Contours of constant $c$ on the $\alpha$, $N^2$ plane.  The Richardson criterion states that the shaded region is stable.  The dotted line represents the $c=0$ contour --- the region below this line is unstable.}\label{fig:cvsN2alpha}
\end{figure}

For this problem, the unstable area of the $(N^2,\alpha$) plane is
maximized for $k^2=0$ --- though this is not necessarily true in
general (K68).  When $k^2=0$ we have $c=c(N^2,\alpha)$.  We have plotted
contours of constant $c$ on the $N^2, \alpha$ plane in Figure
\ref{fig:cvsN2alpha}.

We find instability when $N^2>1/4$, violating the Richardson
criterion.  It seems that given an arbitrarily large value of $N^2$,
there is a sufficiently negative value of $\alpha$ such that the
fields are unstable.  However, as mentioned in \S\ref{sec:TDNneq0}, the
extremely negative $\alpha$ case is probably strongly affected by the
boundary conditions.

Gravity is stabilizing: the growth rate decreases as $N^2$ increases.
There is stability for $\alpha<0.5$ by the same arguments as above,
and as $\alpha$ becomes more negative, we find larger $c$.  Although a
stronger magnetic field results in a strong magnetic tension force,
and the ``destabilizing'' region in which $|V|>|A|$ shrinks for more
negative $\alpha$, we nevertheless find stronger instability.  We
hypothesize that $c$ increases because there is more free energy in
the magnetic field as $\alpha$ becomes more negative and the magnetic
field becomes stronger.  As $\alpha$ becomes more negative, the
instability can tap more free energy from the magnetic field, and thus
we find a violation of the Richardson criterion.  However, note that
the stronger magnetic field, and corresponding increase in magnetic
free energy, is not a sufficient condition for instability, as the
magnetic field is stable without the presence of shear flow.

The contours of constant $c$ are well fit by straight lines.  The
equation for the boundary between the stable and unstable regimes is
\begin{equation}\label{eq:TDstability}
\alpha=0.5-2.65 N^2.
\end{equation}
Thus, for $\alpha<-0.1625$, the Richardson criterion is violated.  The
slopes of the contours become steeper as $c$ increases.  Although
there is instability with arbitrarily large $c$, this does not mean
the instability has arbitrarily large growth rate.  As mentioned in
\S \ref{sec:basiceqn}, the growth rate is formally zero at
$k^2=0$.  Thus, to find the growth rate, we need to understand the
instability at $k^2\neq 0$.

\subsubsection{$k^2 > 0$}\label{sec:TDkneq0}

\begin{figure}[!h]
\plotone{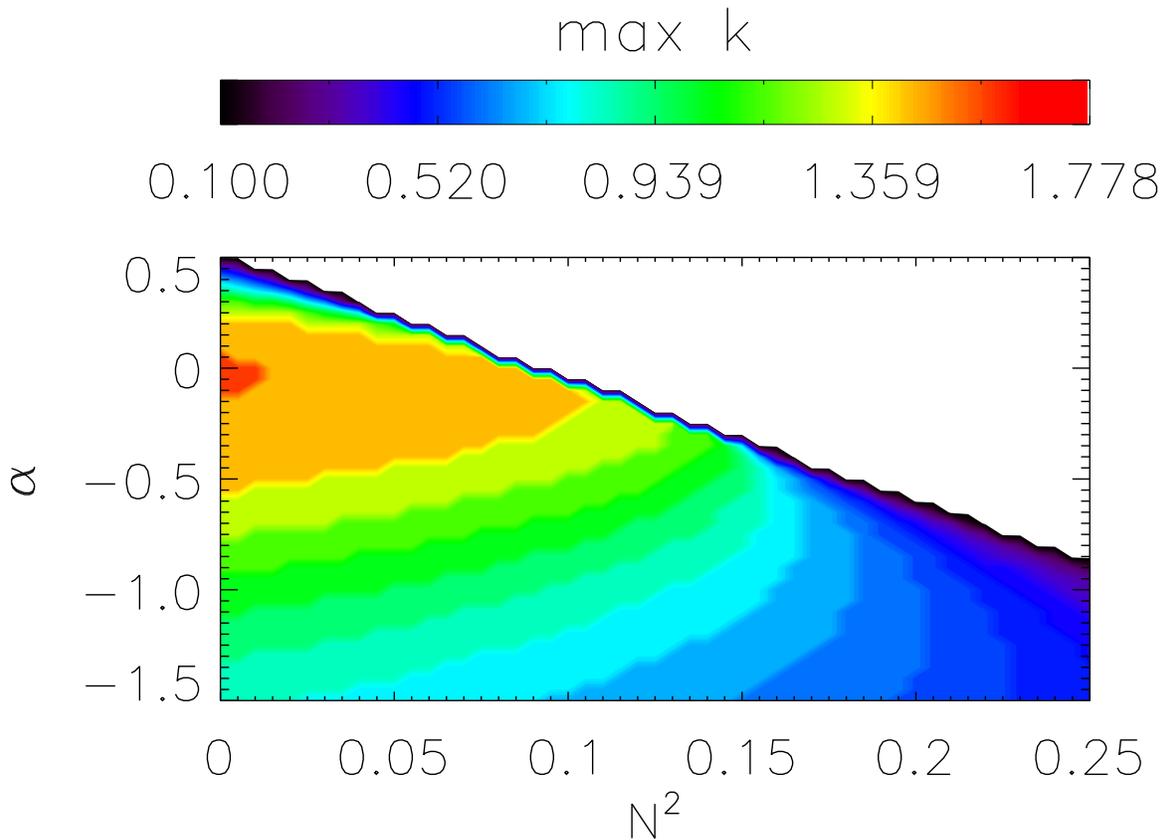}
\caption{The largest $k$, denoted $k_{crit}$,
for each $\alpha$ and $N^2$ which is unstable.  The white area is stable.}\label{fig:maxstablek}
\end{figure}

Although when $k^2=0$ there is instability for arbitrarily negative
$\alpha$, for every finite $k$, there is a cutoff $\alpha_k$ for which
any $\alpha$ more negative than $\alpha_k$ yields stable profiles due to an insurmountable magnetic tension force.
Looking at it another way, $c$ always decreases as $k$ increases, so
for any values $\alpha$ and $N^2$ which are unstable at $k^2=0$, there
is a $k$ for which $c=0$.  Call this value $k_{crit}(\alpha,N^2)$.  Figure
\ref{fig:maxstablek} plots $k_{crit}(\alpha,N^2)$ as a
function of $\alpha$ and $N^2$.  The point $(\alpha,N^2,k)$ is unstable iff $k<k_{crit}(\alpha,N^2)$.  Although it is
possible to find instability when $k^2>0$ for profiles which are
stable when $k^2=0$ (see \S \ref{sec:magnetoshear}), this does
not seem to occur for these classes of profiles.

\begin{figure}[!h]
\plotone{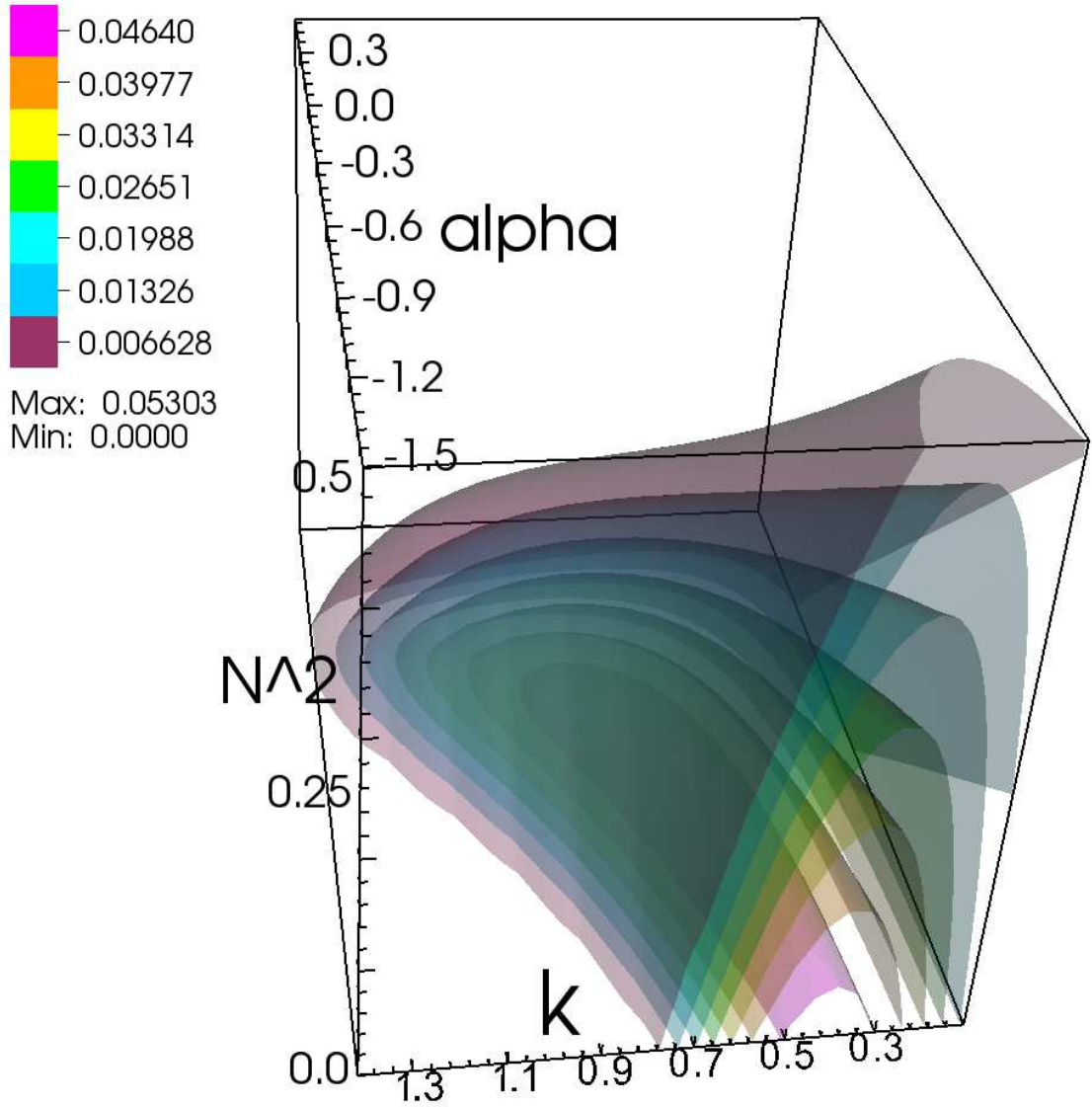}
\caption{Surfaces of constant growth rate $\omega$ in $(\alpha, N^2, k)$ space. The maximum
$\omega$ in this range of $(\alpha, N^2, k)$ is given also.}\label{fig:visit0025}
\end{figure}

Figure \ref{fig:visit0025} plots surfaces of constant $\omega$ in $(\alpha, N^2, k)$
space. The figure shows that $\omega$ is a sharply peaked function of $k$, and that
it decreases with increasing $N^2$.  Given $N^2,k\neq 0$, there is instability for only a finite range of $\alpha$. For
$N^2\equiv 0$, our results
agree with \citet{td06}.
For sufficiently
small $k$, $c$ is almost constant.  Thus, the growth rate
$\omega\equiv kc$ is linear in $k$ with slope $c$.  However, as $k$
grows, $c$ begins to decrease.  There is a maximum growth rate defined by
$d\log c/d\log k=-1$, and the growth rate goes to zero when $c$ does.
The growth rate is 1 -- 2 orders of magnitude lower than the typical growth rates
of hydrodynamic shear flow instabilities.

As $k^2$ increases from zero, the fluid displacement becomes more
vertical.  Vertical perturbations bend field lines, and are subject to
a restoring magnetic tension force.  Thus, it makes sense that the
most unstable modes are the horizontal modes characterized by $k^2=0$.
For some applications, such as stellar interiors (see \S
\ref{sec:stellar}), it is important to consider the vertical transport
(of angular momentum, etc.) by these modes.  In this case, the $k^2=0$
mode is irrelevant.  One must then consider an optimization problem in
which modes with too low $k^2$ have no vertical transport effects,
whereas modes with too high $k^2$ are stable.  This argument is only
valid assuming that the non-linear evolution is similar over a broad
range of $k^2$.  A full non-linear simulation for various $k^2$ is
necessary in order to understand the transport properties of these
instabilities.

\subsection{Eigenfunctions}\label{sec:TDeigenfunction}

\begin{figure}[!h]
\plottwo{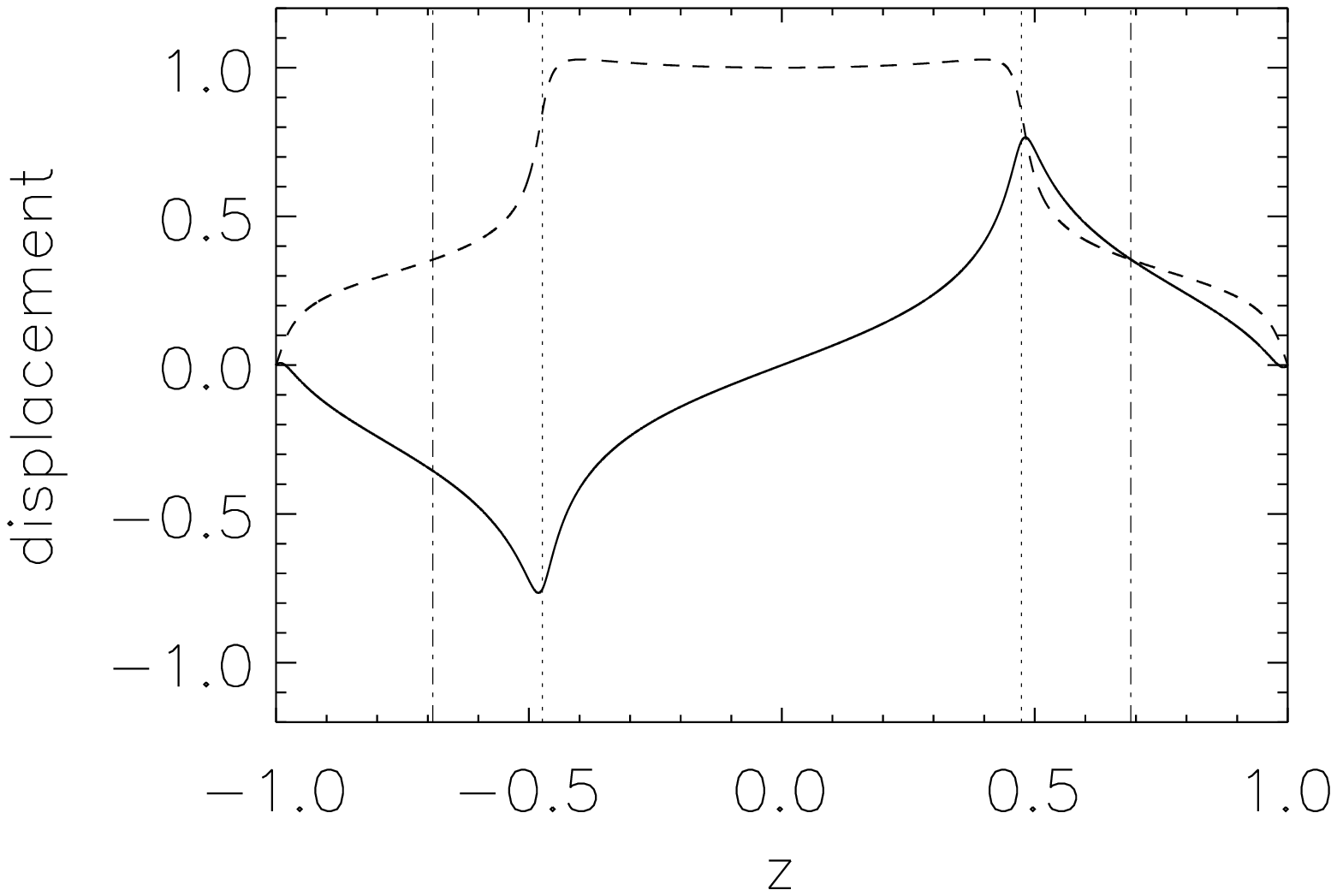}{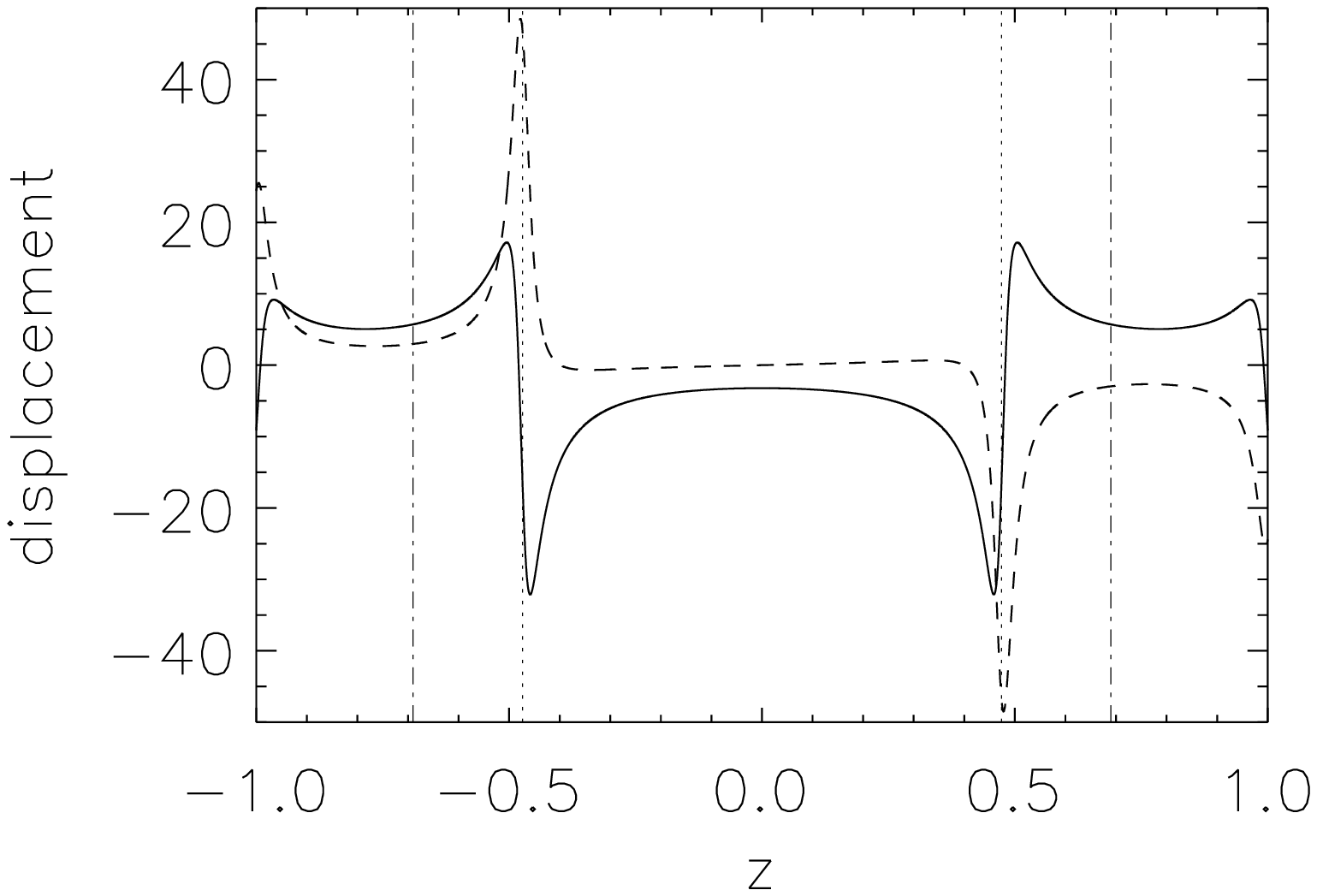}
\caption{The vertical displacement $\xi$ (left panel) and horizontal displacement $\xi_x= -i\xi'/k$ (right panel), where prime denotes differentiation with respect to $z$, eigenfunctions for $\alpha=-0.9$, $N^2=0.3$ and $k=0.2$.  The thick solid lines are the real part of the eigenfunctions, and the thick dashed lines are the imaginary part of the eigenfunctions.  The thin vertical dotted lines are at $z=\pm .47$ where $|V|=|A|$ and the thin vertical dotdashed lines are at $z=\pm .69$, where $A=0$.}\label{fig:TDeigenfunction}
\end{figure}

We normalize the eigenfunctions as described in \S
\ref{sec:basiceqn}.  The eigenfunctions all look like the example
plotted in Figure \ref{fig:TDeigenfunction}.  The most salient
features are the sharp gradients at $z=\pm .47$, where $|V|=|A|$.
Notice that the nulls in the magnetic field at $a=\pm .69$ produce no
special features.

\section{Constant $A$ with Velocity Profiles Suggested by Kent}\label{sec:kent}

In \S\ref{sec:magnetoshear} we summarized Kent's discussion (K68) of
velocity profiles which are marginally stable in the absence of a
magnetic field and destabilized by a small, constant field. In this
section we generalize Kent's construction and investigate the
stability of the resulting family of Kent flows.

The velocity profile is most conveniently specified by the inverse
relation $z=z(V)$.  Note that only invertible velocity profiles, i.e., $dV/dz\neq 0$, can be specified by this inverse relation.  When $k^2=0$ and $N^2=0$, we can use the
instability condition by \citet{cm91} and evaluate the integral in
eqn. (\ref{eq:CMcondition}) in closed form.  This provides a
transcendental equation for the growth rate.  From solving this
equation numerically, it seems that there exist velocity profiles
which are (marginally) stable at $A_0=0$, but unstable for $0< A_0 <
\vert V\vert_{max}$.  When we increase $N^2$ from zero, we always find
stability when $N^2 \geq(\max V')^2/4$, but can find instability for
all $N^2$ up to this limit. Our interpretation of this result is that
the positive energy required to perturb a constant magnetic field
triumphs over the extra freedom granted by magnetically breaking the frozen-in
vorticity constraint.

\subsection{$N^2=0$}\label{sec:kentN=0}

First we consider various velocity profiles defined by $z=z(V)$ at $k^2=0$.  Define
\begin{equation}\label{eq:omega}
\omega(V)\equiv \frac{dz}{dV}.
\end{equation}
We restrict ourselves to velocity profiles which are marginally stable
at $A=0$, as they seem to be maximally destabilized by magnetic
fields.  We will first consider velocity profiles with walls at $z=\pm z_0$, with the condition that $V(\pm z_0)=\pm1$.  This will simplify the algebra when deriving analytic stability results.  We will then employ the rescaling symmetry described in eqn. (\ref{eq:rescale}) to present numerical results using the normalization $z_0=1$.

The condition for marginal stability \citep{kent68} is
\begin{equation}\label{eq:marginalstability}
\int_{-1}^{1}\frac{\omega(V)dV}{V}=0,
\end{equation}
where we have assumed that $V$ ranges from $-1$ to $+1$ in the domain.  Assuming
\begin{equation}\label{eq:z}
z=V+a_3V^3+a_5V^5+\ldots,
\end{equation}
we have
\begin{equation}\label{eq:omegaeven}
\omega=1+3a_3V^2+5a_5V^4+\ldots,
\end{equation}
so the marginal stability condition on the $a_j$'s is
\begin{equation}\label{eq:ms2}
\displaystyle\sum_{j\geq 3,\mbox{ odd}} \frac{ja_j}{j-2}=1.
\end{equation}
Our construction is a generalization of K68, who truncated the series in eqn.
(\ref{eq:z}) at 3 terms.  Next we assume there is only one inflexion
point, at $z=0$.  This condition implies that $\omega$ cannot have any
extrema, so none of the $a_j$ are negative. Numerical work
suggests that the results discussed here hold for velocity profiles
with multiple inflexion points, so by assuming only one inflexion
point, we make the problem much easier, but do not qualitatively
change the results.

Now we add a constant magnetic field.  When $k^2=0$, we have that
\begin{equation}\label{eq:candm}
\int_{-z_0}^{z_0} \frac{dz}{(V-c)^2-A_0^2}=0
\end{equation}
implies instability with growth rate $c$.  If we change variables to $V$, we find
\begin{equation}\label{eq:intcond}
\int_{-1}^{1} \frac{\omega(V) dV}{(V-c)^2-A_0^2}=0,
\end{equation}
where $\omega(V)$ is defined as in eqn. (\ref{eq:omega}).  We can rewrite the integral in
eqn. (\ref{eq:intcond}) as
\begin{equation}\label{eq:intcond2}
\int_{-1}^1 \frac{1}{2A_0} \omega(V) dV \left(\frac{1}{V-c-A_0}-\frac{1}{V-c+A_0}\right)=0.
\end{equation}
The two integrals have equal real parts, so all we need to calculate is
\begin{equation}\label{eq:intcond3}
\Re\int_{-1}^1 \frac{\omega(V) dV}{V-c-A_0}=0.
\end{equation}
When specifying $\omega(V)$ as a power series in odd powers of $V$, as
in eqn. (\ref{eq:omegaeven}), we can evaluate the integral by noticing
that
\begin{eqnarray}\label{eq:intstability}
\frac{1}{2}\int_{-1}^1 \frac{V^n dV}{V-c-A_0}&=&\frac{c+A_0}{n-1}+\frac{(c+A_0)^3}{n-3}+\cdots+(c+A_0)^{n-1} \nonumber \\ &+&\frac{1}{2}(c+A_0)^n\left(\log(1-c-A_0)-\log(-1-c-A_0)\right),
\end{eqnarray}
and summing over each term in the power series for $\omega(V)$.  This
gives a transcendental condition for stability, instead of the
differential condition of eqn. (\ref{eq:eigeqn}).

Notice that the location of the walls plays a crucial role in the equation for stability, eqn. (\ref{eq:intstability}).  Moving the walls from the $z_0$ where $V(z_0)=1$ could make the marginally stable velocity profiles stable or unstable.  Although we will only consider velocity profiles which are marginally stable with no magnetic field below, our results do not change qualitatively when we add a constant magnetic field to a velocity profile which is stable or unstable when $A_0=0$.  We choose marginally stable velocity profiles because they are more clearly destabilized by magnetic fields than unstable velocity profiles, and they are more destabilized than stable velocity profiles.

\begin{figure}[!h]
\plotone{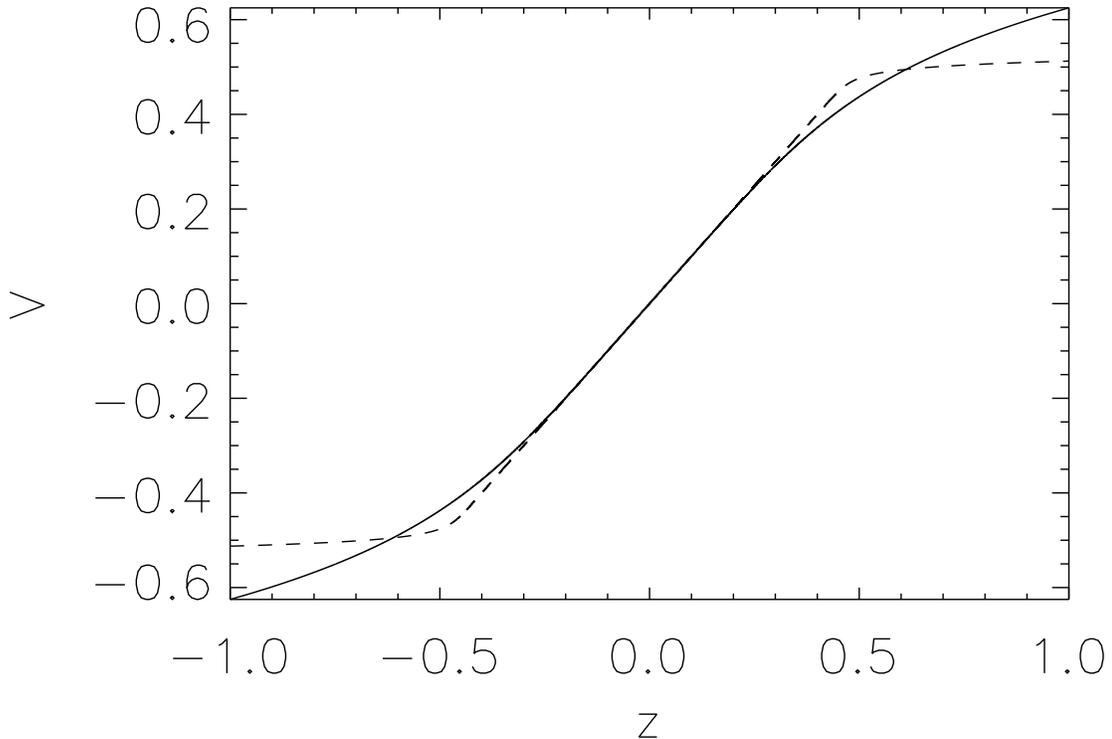}
\caption{The velocity profile solutions of eqn. (\ref{eq:kentvel}) for $n=5$ (solid) and $n=41$ (dashed).}\label{fig:kentprofiles}
\end{figure}

For the remainder of this paper, we will normalize the problem by setting the walls at $z=\pm 1$. Under the assumptions that $V$ has only a single inflexion point and
is marginally stable at $A=0$, we numerically find that the most
unstable velocity profile at $k^2=0$ and $N^2=0$ is given by
\begin{equation}\label{eq:kentvel}
z=V+\left(1+\frac{n-2}{n}\right)^{n-1}\frac{(n-2)V^n}{n},
\end{equation}
for $n$ odd, when $n\rightarrow\infty$. In this limit, the velocity
profile approaches
\begin{equation}\label{eq:nrightarrowinfty}
V(z)=\left\{
\begin{array}{l l}
+\frac{1}{2}, & \frac{1}{2}<z<1 \\
z, & -\frac{1}{2}<z<\frac{1}{2} \\
-\frac{1}{2}, & -1<z<-\frac{1}{2}
\end{array}\right.
\end{equation}
For every $n$ odd and greater than three, the
velocity in eqn. (\ref{eq:kentvel}) is marginally stable.  We plot the velocity profile
for $n=5$ and $n=41$ in Figure \ref{fig:kentprofiles}.  Notice that
$\max V'=1$, so the Richardson criterion states that $N^2>1/4$ yields
stability.

\begin{figure}[!h]
\plotone{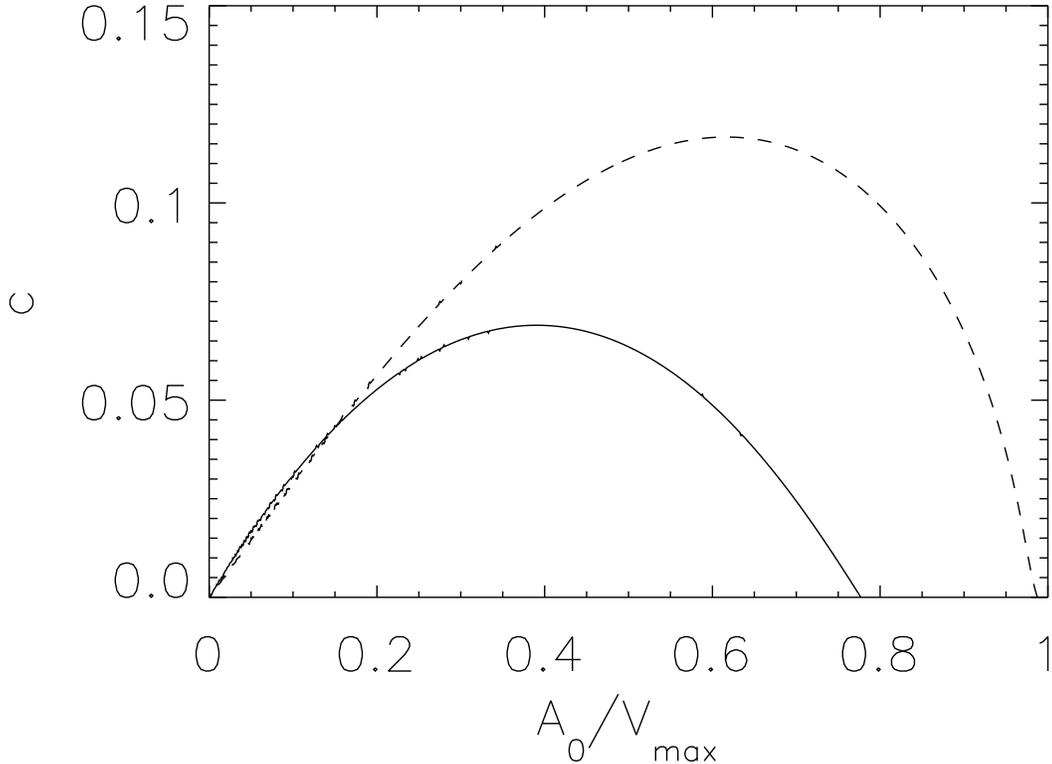}
\caption{The imaginary part of the eigenvalue $c$ as a function of $A_0/V_{max}$ for the velocity profiles given by eqn. (\ref{eq:kentvel}) for $n=5$ (solid) and $n=41$ (dashed).}\label{fig:cvsA}
\end{figure}

For each $n$, we can plot $c$ as a function of $A_0$ at $k^2=0$.  Because we assumed the magnetic field is parallel to the velocity, we know there is stability when $A_0>V_{max}$.  Thus, $V_{max}$ sets a natural scale for measuring the magnetic field strength.
Figure \ref{fig:cvsA} plots $c(A_0/V_{max})$ for $n=5$ and $n=41$.  It seems
that as $n\rightarrow\infty$, the maximum $c$ approaches $\approx
0.125$ for $A_0\approx 0.65 V_{max}=0.325$.

\begin{figure}[!h]
\plotone{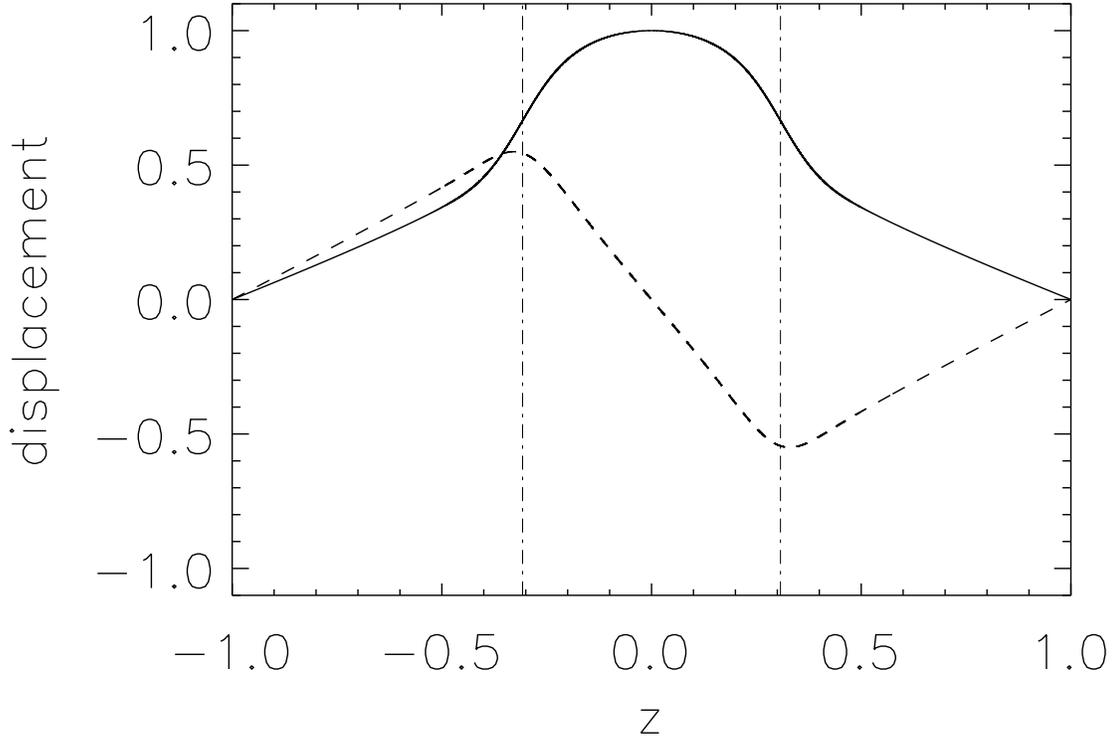}
\caption{Eigenfunction for $A=0.6V_{max}\approx 0.31$ and velocity given by
  eqn. (\ref{eq:kentvel}) for $n=41$.  The thick solid line is the real
  part of the eigenfunction, and the thick dashed line is the
  imaginary part of the eigenfunction.  The vertical dotted lines
  denote the points where $|V|=|A|$, at $z=\pm
  0.40$.}\label{fig:kenteigenfunction}
\end{figure}

Figure \ref{fig:kenteigenfunction} shows an eigenfunction for
$A_0=0.65V_{max}\approx 0.31$, $n=41$.  Note that it is very similar to the eigenfunction
for the Tatsuno \& Dorland profiles in \S
\ref{sec:TDeigenfunction}.

\pagebreak

\subsection{$N^2\neq 0$}\label{sec:kentNneq0}

\begin{figure}[!h]
\plotone{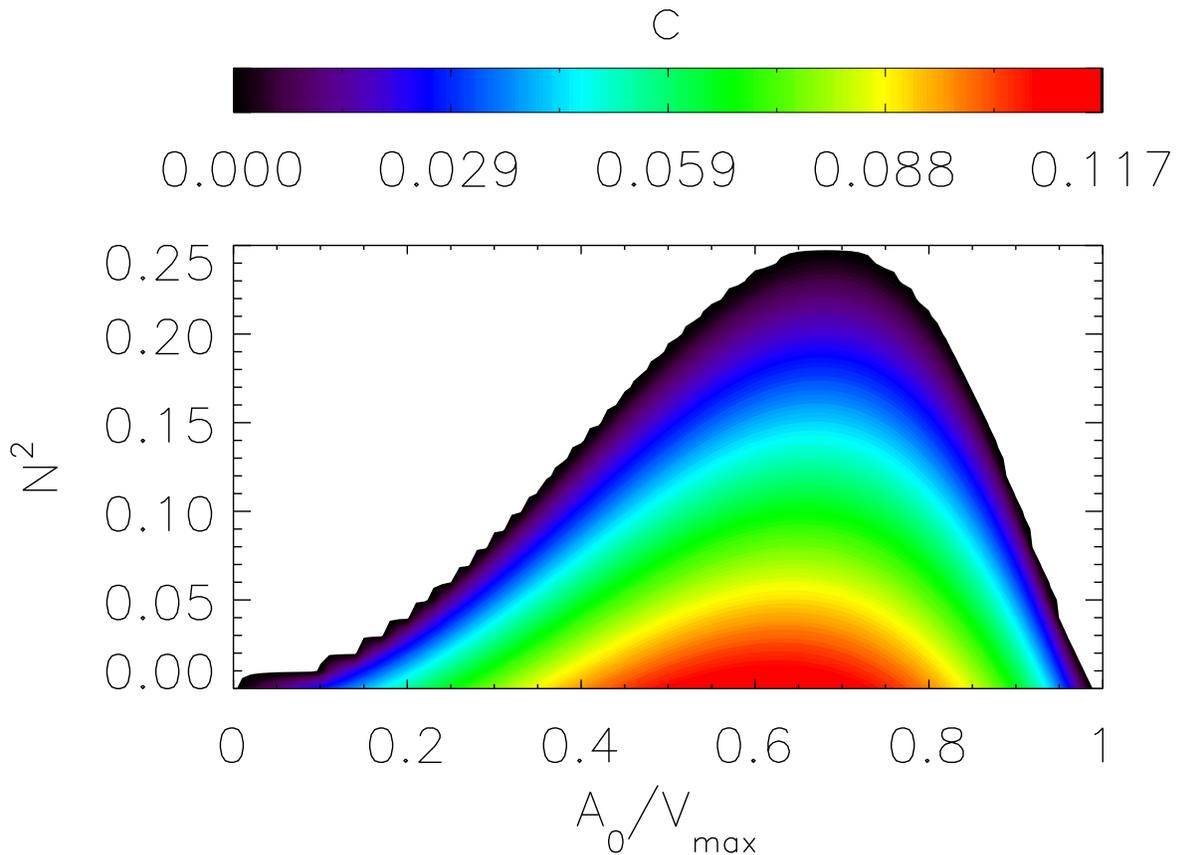}
\caption{Contours of $c$ as a function of $N^2$ and $A_0/V_{max}$ for $k=0$ and the velocity profile given by eqn. (\ref{eq:kentvel}) with $n=41$.  The white area is stable.}\label{fig:cvsN2A0}
\end{figure}

As mentioned in \S \ref{sec:kentN=0}, the velocity profiles
considered here have $\max V'=1$, so the Richardson criterion states
that $N^2>1/4$ implies stability.  As $n$ increases, the maximally
unstable $N^2$ increases, but never seems to reach $1/4$.  Figure
\ref{fig:cvsN2A0} shows contours of $c$ as a function of $N^2$ and
$A_0/V_{max}$ for $k=0$ and $n=41$.  Although there is instability for $N^2$
very close to $1/4$, we find stability at $N^2=0.25$.  It seems that
the Richardson criterion is not violated when adding a constant
magnetic field to this class of velocity profiles.

\begin{figure}[!h]
\plotone{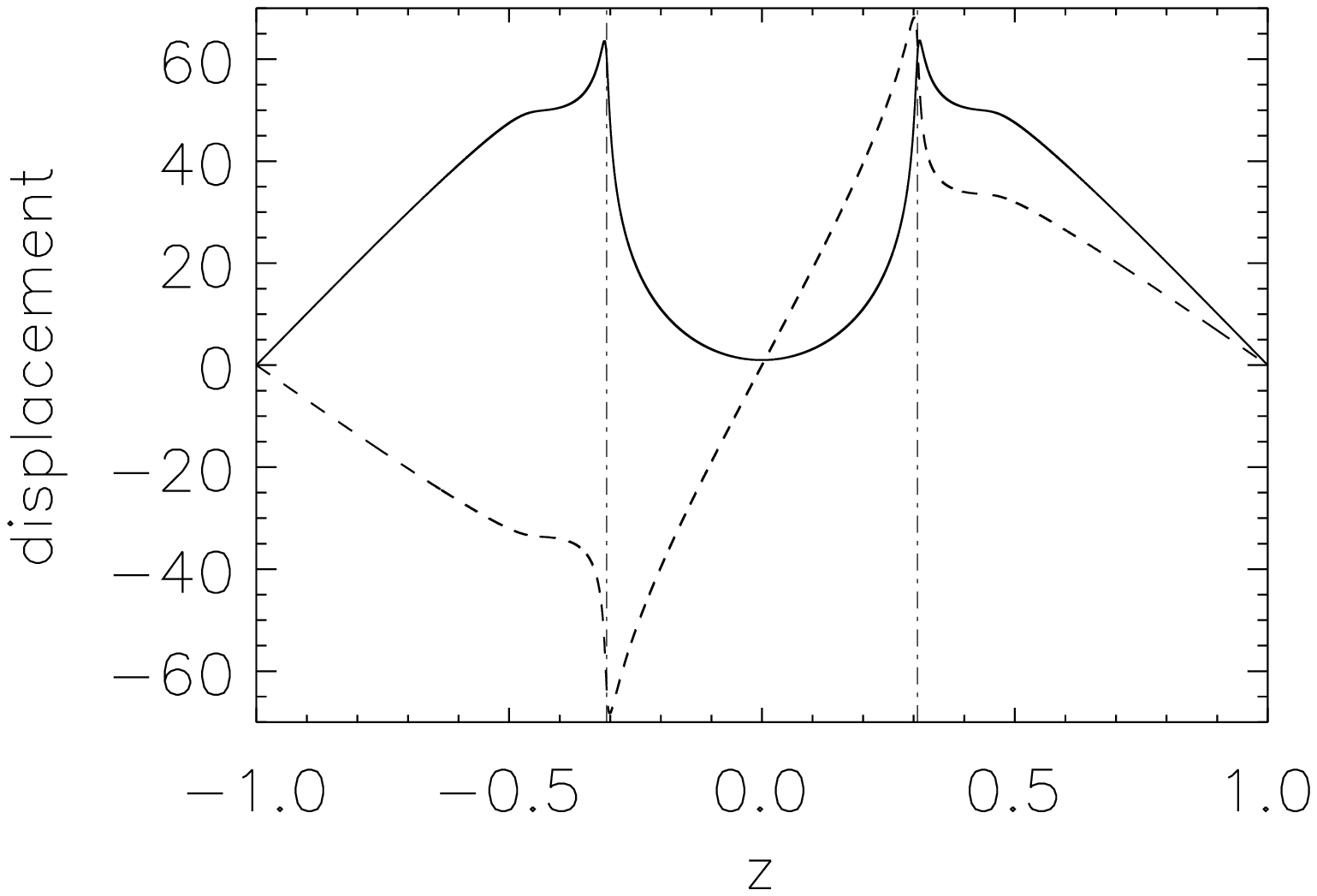}
\caption{Eigenfunction for $A=0.6V_{max}\approx 0.31$ and velocity given by eqn. (\ref{eq:kentvel}) for $n=41$, with $N^2=0.225$, $k=0$.  The thick solid line is the real part of the eigenfunction, and the thick dotted line is the imaginary part of the eigenfunction.  The vertical dotdashed lines are at $z\approx\pm 0.31$, where $|V|=|A|$.}\label{fig:kenteigenfuncNneq0}
\end{figure}

Figure \ref{fig:kenteigenfuncNneq0} shows a typical eigenfunction.  As
with the velocity and magnetic field profiles considered in \S
\ref{sec:TD}, there are sharp gradients when $|V|=|A|$.  Unlike the
eigenfunctions considered above, the real part of this eigenfunction
is close to zero at the origin.

The constant magnetic field case is very different from the parabolic
case because there is no violation of the Richardson Criterion.  We
can understand result heuristically by noting that a constant magnetic
field cannot increase the free energy of the perturbation, and thus
cannot render a velocity profile with $N^2>\max V'^2/4$ unstable.  Although there is no energy principle in the presence of shear flow, one can show that a sufficient condition for stability is that the energy of a perturbation is positive, i.e. $\mathbf{F}(\mbox{\boldmath$\xi$}) \cdot\mbox{\boldmath$\xi$}>0$, where $\mathbf{F}(\mbox{\boldmath$\xi$}) $ is the force operator \citep{fr60}.  A constant magnetic
field contributes $+|\mathbf{Q}|^2$ to the energy of a
perturbation, where $\mathbf{Q}=\mbox{\boldmath$\nabla$}\times
(\mbox{\boldmath$\xi$}\times\mathbf{B})$.  Thus, a constant magnetic
field always increases the energy of a perturbation.

However, in \S \ref{sec:kentN=0}, we describe an entire class of
velocity profiles which are (marginally) stable at $A_0=0$, but
unstable for $A_0>0$.  Our interpretation of the destabilized is as follows.  An unstable perturbation must have negative
energy \citep{fr60}, but this is only a necessary condition for
instability.  Thus, perturbations to the velocity profiles considered in \S
\ref{sec:kentN=0} have negative energy, but are still stable.  For a
sufficiently small magnetic field ($A_0<V_{max}$), the
increase in energy of the perturbation from the magnetic field can
be overcome by a negative contribution from the shear flow, so the
total energy of the perturbation is negative and there could be
instability.

Because the Richardson criterion can be understood from energetic
arguments (see \S \ref{sec:shearstrat}), one could assume that
when $N^2>V'^2/4$ in the entire domain that the energy is necessarily
positive.  Then the addition of a constant magnetic field only further
increase the energy of the perturbation, preventing instability.  This is a rather
considerable assumption, so this argument is best viewed as a
heuristic.

\section{Application to Astrophysical Systems}\label{sec:stellar}

We have studied shear flow instability in stably stratified media for flow profiles which would be
stable in the absence of a magnetic field and shown that 
Richardson's criterion
for buoyancy stabilization can be violated, provided that the magnetic field is inhomogeneous. In this
section we briefly discuss astrophysical applications.

First, some general considerations. Our analysis holds when the flow and field are perpendicular to gravity. We 
ignored the effect of the magnetic field on the density stratification, thereby precluding any instabilities
associated with magnetic buoyancy. Thus, our work applies primarily to situations
in which the field is not too strong and its scale height is not much less than the pressure scale height.
Thus, although we gave an example in \S\ref{sec:TD} of a system that can be unstable at
arbitrarily large $\mbox{Ri}$, instability at large
$\mbox{Ri}$ required in that case that the flow be 
sub-Alfv\'{e}nic in most of the domain and that
the magnetic scale length be much less than the velocity
shear length. In addition to the possible introduction of magnetic buoyancy effects, a small magnetic scale height
relative to the velocity scale height requires that the magnetic Prandtl number $\mbox{Pm}$ --- the ratio of viscous to magnetic
diffusivity --- be much greater than unity, opposite to the situation in dense plasmas such as
stellar interiors. Bearing these things in mind, there is probably a practical upper limit on $\mbox{Ri}$ at which
magnetic fields are destabilizing according to the mechanism discussed here.

It is useful to cast $\mbox{Ri}$ in a form which allows its magnitude to be estimated. We introduce a buoyancy parameter
$f_{bu}$ in terms of which $N^2$ can be written in terms of the local gravity and pressure scale height as 
\begin{equation}\label{N2par}
N^2=f_{bu}\frac{g}{H_{\rho}},
\end{equation}
where $g$ and $H_{\rho}$ are the local gravity and density scale height, respectively; in the Boussinesq approximation,
$f_{bu}=1$. Specializing to the case that $V$ is a rotational velocity, we introduce the velocity scale height $H_v$ by
$V^{'}=V/H_v$ and a breakup parameter $f_{br}$ by
\begin{equation}\label{V2par}
\vert V^{'}\vert^2=f_{br}\frac{rg}{H_v^2},
\end{equation}
where $r$ is the distance from the rotation axis.  Using eqns. (\ref{N2par}) and (\ref{V2par}), $\mbox{Ri}$ can be written as
\begin{equation}\label{Ripar}
\mbox{Ri}=\frac{f_{bu}}{f_{br}}\frac{H_v}{H_{\rho}}\frac{H_v}{r}.
\end{equation}

In stably stratified systems with uniform composition, $f_{bu}$ is generally ${\mathcal{O}}(1)$, while a molecular weight
gradient can render $f_{bu}\gg 1$. Except for systems rotating near breakup, $f_{br}\ll 1$. Typically, $H_v$ exceeds
the geometric width of a shear layer because $V$ changes by only a fraction of itself. Thus, although the second and third
ratios on the right hand side of eqn. (\ref{Ripar}) are below unity, they are generally not enough to offset $f_{bu}/f_{br}$, and $\mbox{Ri}\gg 1$.
One exception to these considerations occurs near the boundaries of convection zones, where $N^2$ crosses through zero. Thus, a thin layer
on the stably stratified side of the boundary could be magnetically destabilized even if $\mbox{Ri} > 1/4$.

The expectation that $\mbox{Ri}\gg 1$ in the stably stratified portions of stellar interiors is borne out by examination of stellar models. First, we consider
the Sun. Helioseismology has revealed a thin
shear layer, known as the tachocline, below the base of the solar convection zone, which is thought to lie at $0.713 R_{\odot}$ (see \citep{Gou2007} for a
review). If we take $N^2$ at $0.700 R_{\odot}$ from Gough and $V^{'}$ from \citet{Sch2000}, we
find that at the equator $\mbox{Ri}=6400$ and $f_{bu}\sim 10^{-2}$. In other words, even very close to
the base of the convection zone  $\mbox{Ri}$ is quite large, and increases with depth from the value
given here.

We also evaluated $\mbox{Ri}$ in an evolutionary sequence of models of massive, rotating stars
generously provided to us by G. Meynet. The initial mass is 20 $M_{\odot}$ (which decreases due
to mass loss) and the initial surface rotation period is about 1.2 $d$. When the star first reaches
the main sequence, the core is
convective and the envelope is radiative. As hydrogen is exhausted in the core, strong nonhomologous
contraction spins up the core and creates strong shear layers, which tends to reduce $\mbox{Ri}$.
At the same time, steep negative molecular weight gradients increase $f_{bu}$. We find that in the bulk
of the interior, $\mbox{Ri}$ is between 10$^2$ and 10$^6$. In the models, the boundaries of convection
zones (which form in association with shell burning) actually show spikes in $\mbox{Ri}$. This is because
$\Omega$ is set to a constant in convection zones, due to efficient turbulent mixing. Thus, although
there is probably a thin layer in which $\mbox{Ri}$ drops to small values, it cannot be evaluated from
these models.

These estimates suggest that destabilization of stellar rotation profiles by weak magnetic fields is 
likely to occur only in thin layers outside convection zones. However, the tendency for such fields to
destabilize a system may be important even when physical processes neglected by our analysis are
included. Chief among them is thermal diffusion, which can suppress the stabilizing effects of buoyancy
\citep{Zah1974} and leads to a larger critical $\mbox{Ri}$ to guarantee stabilization. Whether this carries
over our analysis is a topic for future study.
 
The instability could conceivably also operate on poloidal flows. However, because such flows are generally
slow compared with rotation, their $\mbox{Ri}$ tends to be even larger than $\mbox{Ri}$ for rotation.
And because rotational shear tends to make the magnetic field predominantly toroidal, magnetic effects on the stability of poloidal flow are probably weak.

Similar considerations hold for accretion disks. The vertical shear in a Keplerian disk of thickness $H$ is smaller than the
radial shear by a factor of $H/r$. If the radial inflow velocity is a function of height, its shear could be large,
but the magnetic field is expected to be predominantly toroidal. 
Therefore, this instability is probably not critically important for either rotation or radial flow in disks.

\section{Conclusion}\label{sec:conclusion}

Turbulence is a key ingredient in the transport of chemical species,
entropy, angular momentum, and magnetic flux in astrophysical
settings. Shear flows, which are driven almost ubiquitously in nature,
can become turbulent through instability.

In this paper we have considered ideal instabilities of magnetized
shear flows in stably stratified systems. In the absence of magnetic
fields, the Richardson criterion provides a necessary condition for
instability based on comparing the kinetic energy released by vertical
interchange of fluid elements to the potential energy required to
displace them. The Richardson criterion is often assumed to set the
ideal stability boundary for shear flow instabilities in stratified
media such as stars and accretion disks.  The main result of this
paper is that the Richardson criterion is no longer valid when
inhomogeneous magnetic fields are included: because such fields carry
free energy, buoyancy forces must be stronger to stabilize the system.
We have provided an example by adding density stratification to the
fields described by \citet{td06}.  These fields can be viewed as a
local approximation of any shear flow in the presence of a magnetic
extremum. The system has the interesting property that the flow is
neutrally stable in the absence of the magnetic field, but unstable in
its presence.  Solving the eigenvalue problem in
eqn. (\ref{eq:eigeqn}), we find unstable modes for arbitrarily large
$N^2$, provided the magnetic field is sufficiently strong. Even for
magnetic fields yielding Alfv\'{e}n velocities comparable to flow
velocities, we find violation of the Richardson criterion.  Thus, when
considering the ideal stability of a plasma shear flow in a stratified
medium, it is not sufficient to consider the Richardson criterion.

We were unable to find an example in which a {\textit{constant}}
magnetic field leads to violation of the Richardson criterion. We
extended and analyzed a class of velocity profiles considered by
\citet{kent68}, which were shown to be destabilized by a constant
magnetic field.  Although we were able to destabilize the flows when
$N^2=0$, and the fastest growing modes have moderately strong magnetic
fields, when $N^2>V'^2/4 $, we always found stability.
We provided two heuristics for understanding the destabilization due to 
magnetic fields.  An inhomogeneous magnetic field provides a free energy 
source which can be tapped by an instability.  Thus, while a homogeneous 
magnetic field can be destabilizing because vorticity is no longer frozen 
into the flow, allowing new unstable plasma motions, only an inhomogeneous field can provide the source of energy needed to violate Richardson's criterion.

We briefly applied our results to the solar tachocline and to high mass, rapidly
rotating stars. In the bulk of the tachocline, $\mbox{Ri}$ is very large because
the Sun rotates slowly. Very near the boundary of the convection zone, $\mbox{Ri}$
drops because $N^2$ is passing through zero. A similar situation holds, for different
reason, in high mass stars. Although these stars rotate rapidly, the regions of
strong shear coincide with regions of strong, stabilizing, molecular weight gradient.
This keeps $\mbox{Ri}$ large, except near convection zone boundaries.
Thus, in stars, the destabilization of stratified shear flow by magnetic fields is most
likely to occur in thin regions on the stable side of convection zone boundaries.  If
the weakening of buoyancy by thermal diffusion destabilizes magnetized flow in the same
way as unmagnetized flow, the unstable region could be much larger, however.

Our 2D slab model is not a realistic geometry for many applications.
The introduction of additional terms, such as curvature terms from
toroidal geometry or the centrifugal force for rotation, probably
changes our results quantitatively, but not qualitatively.  The
Boussinesq approximation could also be relaxed to allow more realistic
density profiles and other physics. Inclusion of diffusive effects
would allow us to consider non-ideal instabilities, including the
secular shear instability.  For many applications, the non-linear
phase and saturation of these instabilities is also important for
determining effects such as angular momentum transport.  These
considerations should be investigated further to better understand the
nature of magneto-shear instabilities in a stratified medium.

\acknowledgments

This work was supported by the University of Wisconsin -- Madison
Hilldale Undergraduate/Faculty Research Fellowship to DL and EGZ, NSF
Cooperative Agreement PHY-0821899 which funds the Center for Magnetic
Self-Organization, NSF Grants AST-0507367 and AST-0903900, NASA Grant
LTSA NNG05GC36G,  and the University of Wisconsin -- Madison Graduate School. We are happy to
acknowledge useful discussions with B. Brown, F. Ebrahimi, J. Everett, \& I. Shafer, and grateful to
G. Meynet for supplying us with models of massive, rotating stars. 

\appendix

\section{Instability of $V=z$, $A=(1-\alpha)z^2+\alpha$ when $\alpha<0.5$}\label{sec:alphainf}

We will prove that the velocity and magnetic field profiles considered
in \S \ref{sec:TD}, $V=z$, $A=(1-\alpha)z^2+\alpha$, are unstable
when $\alpha<0.5$.  In \S \ref{sec:magnetoshear} we described the
following sufficient condition for instability at $k^2=0$ by Chen and
Morrison (eqn. (\ref{eq:CMcondition})): If
\begin{equation}\label{eq:CMalpha}
\int_{-1}^{1} \frac{1}{(V-i\epsilon)^2-A^2}>0
\end{equation}
as $\epsilon\rightarrow 0$, then there is instability.  We can factor
the denominator to get
\begin{equation}\label{eq:alphafactor}
\frac{1}{2}\int_{-1}^{+1}\frac{dz}{A(V-i\epsilon-A)}-\frac{1}{2}\int_{-1}^{+1}\frac{dz}{A(V-i\epsilon+A)}.
\end{equation}
Let us examine how these two integrals are related.  Define $u=-z$.  Then
\begin{eqnarray}\label{eq:alphasecondint}
-\frac{1}{2}\int_{-1}^{+1}\frac{dz}{A(z)(V(z)-i\epsilon+A(z))}&=&\frac{1}{2}\int_{+1}^{-1}\frac{du}{A(z)(V(z)-i\epsilon+A(z))} \\
&=&-\frac{1}{2}\int_{-1}^{+1}\frac{du}{A(u)(-V(u)-i\epsilon+A(u))}\nonumber \\
&=&\frac{1}{2}\int_{-1}^{+1}\frac{du}{A(u)(V(u)+i\epsilon-A(u))}\nonumber,
\end{eqnarray}
which has the same real part as the first integral, but opposite
imaginary part.  Thus, we need only check that
\begin{equation}\label{eq:alphaoneint}
\Re\int_{-1}^{+1}\frac{dz}{A(V-i\epsilon-A)}>0
\end{equation}
as $\epsilon\rightarrow 0$ to prove instability.  Integrals of this
form can be evaluated in closed form, but must first be factored.  To
simplify the algebra, we reduce the degree of the polynomial in the
denominator through partial fractions.
\begin{equation}\label{eq:alphafactor2}
\Re\int_{-1}^{+1}\frac{dz}{A(V-i\epsilon-A)}=\Re\int_{-1}^{+1}\frac{dz}{A(V-i\epsilon)}+\Re\int_{-1}^{+1}\frac{dz}{(V-i\epsilon)(V-i\epsilon-A)}
\end{equation}
The first integral gives no contribution because multiplying by
$V+i\epsilon$ in the numerator and denominator shows that the real
part is odd and integrates to zero.  Thus, we need only evaluate the
second integral.

We can integrate the remaining part by brute force, i.e. using
Mathematica.  Assuming $\epsilon>0$, Mathematica gives
\begin{eqnarray}\label{eq:alphaint}
& &\int\frac{dz}{(V-i\epsilon)(V-i\epsilon-A)}=-\frac{1}{4(\alpha-\epsilon^2+\alpha\epsilon^2)}\left[-4i\arctan\left(\frac{\epsilon}{z}\right)\right. \\
&-&\log\left(\epsilon^2+(-1+z)^2z^2-2\alpha (-1+z)^2z(1+z)+\alpha^2(-1+z^2)^2\right)\nonumber \\
&+&\frac{4(1-2i(1-\alpha)\epsilon)}{\sqrt{-1-4\alpha^2+4i\epsilon-4i\alpha (i+\epsilon)}}\arctan\left(\frac{-1+2(1-\alpha)z}{\sqrt{-1-4\alpha^2+4i\epsilon-4i\alpha (i+\epsilon)}}\right) \nonumber\\
&+&\left. 2i\arctan \left(\frac{(-1+z)(z-\alpha z-\alpha)}{\epsilon}\right)+2\log\left(\epsilon^2+z^2\right)\right].\nonumber
\end{eqnarray}
Notice that the prefactor has the opposite sign as $\alpha$.  The term
on the first line is imaginary, so we do not need to consider it.  In
the logarithm on the second line, the third and fourth terms which are
0 at $z=\pm 1$.  On the last line, the first term is imaginary and the
second term is even, so neither contribute to the integral.  Thus, if
\begin{eqnarray}\label{eq:alphacondition}
&-&\frac{1}{4(\alpha-\epsilon^2+\alpha\epsilon^2)}\left[\log\left(\frac{\epsilon^2+4}{\epsilon^2}\right) \right. \\
&+&\Re \frac{4(1-2i(1-\alpha)\epsilon)}{\sqrt{-1-4\alpha^2+4i\epsilon-4i\alpha (i+\epsilon)}}\arctan\left(\frac{-1+2(1-\alpha)}{\sqrt{-1-4\alpha^2+4i\epsilon-4i\alpha (i+\epsilon)}}\right) \nonumber \\
&-&\left.\Re \frac{4(1-2i(1-\alpha)\epsilon)}{\sqrt{-1-4\alpha^2+4i\epsilon-4i\alpha (i+\epsilon)}}\arctan\left(\frac{-1-2(1-\alpha)}{\sqrt{-1-4\alpha^2+4i\epsilon-4i\alpha (i+\epsilon)}}\right)\right]>0 \nonumber
\end{eqnarray}
for a particular $\alpha$ as $\epsilon\rightarrow 0$, then the
profiles for that $\alpha$ are unstable.  The $\epsilon$ for which the
RHS of eqn. (\ref{eq:alphacondition}) equals zero is the growth rate
of the instability.  Thus, this relation gives a transcendental
equation for the growth rate, which is significantly easier to solve
than the differential eigenvalue problem given in \S
\ref{sec:basiceqn}.

As $\epsilon\rightarrow 0$, the logarithm term diverges and is
positive.  However, when $z=+1$, the $\arctan$ term also diverges,
approaching $-i\infty$, meaning that the entire term gives a negative
divergent contribution.  We need to see which diverges faster.  The
argument of the $z=+1$ $\arctan$ term is
\begin{eqnarray}\label{eq:alphaargarctan}
\frac{1-2\alpha}{\sqrt{-1-4\alpha^2+4i\epsilon-4i\alpha (i+\epsilon)}}&=&-i\frac{1-2\alpha}{\sqrt{4\alpha^2-4\alpha+1-4i\epsilon(1-\alpha)}} \nonumber\\
&=&-i\left(1-\frac{4i\epsilon(1-\alpha)}{4\alpha^2-4\alpha+1}\right)^{-1/2} \nonumber\\
&\approx&-i\left(1+\frac{1}{2}\frac{4i\epsilon(1-\alpha)}{(1-2\alpha)^2}\right).
\end{eqnarray}
In general, $\arctan(z)$ is given by
\begin{equation}\label{eq:alphaarctan}
\arctan(z)=i\frac{1}{2}\left(\log(1-iz)-\log(1+iz)\right).
\end{equation}
The divergent part for us is the first term, so
\begin{equation}\label{eq:alphadivarctan}
\arctan\left(\frac{-1-2(1-\alpha)}{\sqrt{-1-4\alpha^2+4i\epsilon-4i\alpha (i+\epsilon)}}\right)\approx i\frac{1}{2}\log\left(-\frac{2i\epsilon(1-\alpha)}{(1-2\alpha)^2}\right).
\end{equation}
If we neglect the $\epsilon$ terms which are not in the divergence, we
find that the coefficient of the $\log(\epsilon)$ term is
$-2/(1-2\alpha)$.  Thus, only considering the terms in
eqn. (\ref{eq:alphacondition}) which are divergent as
$\epsilon\rightarrow 0$, and taking $\epsilon=0$ except for in the
divergence, we are left with
\begin{equation}\label{eq:alphaconddiv}
-\frac{1}{4\alpha}\left(2\log(\epsilon)-\frac{2}{1-2\alpha}\log(\epsilon)\right).
\end{equation}
When $\alpha<0$, we have that $-1/4\alpha>0$, and the first
$\log(\epsilon)$ term dominates, so the whole quantity is positive.
Thus, we have proven that there is instability for $\alpha<0$.  When
$0.5>\alpha>0$, we have $-1/4\alpha<0$, but the second logarithm term
dominates and is negative, again yielding instability.  However, when
$\alpha>0.5$, both divergent terms become positive, but
$-1/4\alpha<0$, so the quantity is negative as $\epsilon\rightarrow
0$, and the profiles are stable.  In order to show instability at
$\alpha=0$, we would need to retain more terms in our perturbative
expansion in $\epsilon$.

\bibliographystyle{apj}

\bibliography{ssf}

\begin{thebibliography}{40}
\expandafter\ifx\csname natexlab\endcsname\relax\def\natexlab#1{#1}\fi

\bibitem[{{Chandrasekhar}(1961)}]{chandra61}
{Chandrasekhar}, S. 1961, {Hydrodynamic and Hydromagnetic Stability}
  ({Clarendon Press})

\bibitem[{{Chen} \& {Morrison}(1991)}]{cm91}
{Chen}, X.~L., \& {Morrison}, P.~J. 1991, {Phys. Fluids B}, 3, 863

\bibitem[{{Dikpati} {et~al.}(2009){Dikpati}, {Gilman}, {Cally}, \&
  {Miesch}}]{dg09}
{Dikpati}, M., {Gilman}, P.~A., {Cally}, P.~S., \& {Miesch}, M.~S. 2009, \apj,
  692, 1421

\bibitem[{{Drazin} \& {Reid}(1981)}]{dr81}
{Drazin}, P.~G., \& {Reid}, W.~H. 1981, {Hydrodynamic Stability} ({Cambridge U.
  P., London})

\bibitem[{{Frieman} \& {Rotenberg}(1960)}]{fr60}
{Frieman}, E.~A., \& {Rotenberg}, M. 1960, {Rev. Mod. Phys.}, 32, 898

\bibitem[{{Furth} {et~al.}(1963){Furth}, {Killeen}, \& {Rosenbluth}}]{FKR63}
{Furth}, H.~P., {Killeen}, J., \& {Rosenbluth}, M.~N. 1963, {Phys. Fluids}, 6,
  459

\bibitem[{{Gamelin}(2001)}]{gamelin01}
{Gamelin}, T.~W. 2001, {Complex Analysis} ({Springer})

\bibitem[{{Gilman} {et~al.}(2007){Gilman}, {Dikpati}, \& {Miesch}}]{Gil2007}
{Gilman}, P.~A., {Dikpati}, M., \& {Miesch}, M.~S. 2007, \apjs, 170, 203

\bibitem[{{Gilman} \& {Fox}(1997)}]{GilFox1997}
{Gilman}, P.~A., \& {Fox}, P.~A. 1997, \apj, 484, 439

\bibitem[{{Gough}(2007)}]{Gou2007}
{Gough}, D. 2007, in The Solar Tachocline, ed. {D.~W.~Hughes, R.~Rosner, \&
  N.~O.~Weiss}, 3--+

\bibitem[{{Howard}(1961)}]{howard61}
{Howard}, L.~N. 1961, {J. Fluid Mech.}, 10, 509

\bibitem[{{Howes} {et~al.}(2001){Howes}, {Cowley}, \& {McWilliams}}]{h01}
{Howes}, G.~G., {Cowley}, S.~C., \& {McWilliams}, J.~C. 2001, \apj, 560, 617

\bibitem[{{Hughes} \& {Tobias}(2001)}]{ht01}
{Hughes}, D.~W., \& {Tobias}, S.~M. 2001, {Proc. R. Soc. Lond. A}, 457, 1365

\bibitem[{{Kent}(1968)}]{kent68}
{Kent}, A. 1968, {J. Plasma Physics}, 2, 543

\bibitem[{{Keppens} {et~al.}(1999){Keppens}, {T\'{o}th}, {Westermann}, \&
  {Goedbloed}}]{keppens99}
{Keppens}, R., {T\'{o}th}, G., {Westermann}, R.~H.~J., \& {Goedbloed}, J.~P.
  1999, {J. Plasma Physics}, 61, 1

\bibitem[{{Kitchatinov} \& {R{\"u}diger}(2009)}]{KitRud2009}
{Kitchatinov}, L.~L., \& {R{\"u}diger}, G. 2009, \aap, 504, 303

\bibitem[{{Krall} \& {Trivelpiece}(1973)}]{kt73}
{Krall}, N.~A., \& {Trivelpiece}, A.~W. 1973, {Principles of Plasma Physics}
  ({McGraw-Hill})

\bibitem[{{Lin}(1955)}]{lin55}
{Lin}, C.~C. 1955, {The Theory of Hydrodynamic Stability} ({Cambridge U. P.,
  London})

\bibitem[{{Maeder}(1995)}]{Mae1995}
{Maeder}, A. 1995, \aap, 299, 84

\bibitem[{{Maeder} \& {Meynet}(1996)}]{MaeMey1996}
{Maeder}, A., \& {Meynet}, G. 1996, \aap, 313, 140

\bibitem[{{Maeder} \& {Meynet}(2000{\natexlab{a}})}]{MaeMey2000}
---. 2000{\natexlab{a}}, \araa, 38, 143

\bibitem[{{Maeder} \& {Meynet}(2000{\natexlab{b}})}]{MM00}
---. 2000{\natexlab{b}}, \araa, 38, 143

\bibitem[{{Maeder} \& {Meynet}(2004)}]{MaeMey2004}
---. 2004, \aap, 422, 225

\bibitem[{{Meynet} \& {Maeder}(2000)}]{MeyMae2000}
{Meynet}, G., \& {Maeder}, A. 2000, \aap, 361, 101

\bibitem[{{Newcomb}(1961)}]{n61}
{Newcomb}, W.~A. 1961, {Phys. Fluids}, 4, 391

\bibitem[{{Ogilvie}(2007)}]{Ogi2007}
{Ogilvie}, G.~I. 2007, in The Solar Tachocline, ed. {D.~W.~Hughes, R.~Rosner,
  \& N.~O.~Weiss}, 299--+

\bibitem[{{Petrovic} {et~al.}(2005){Petrovic}, {Langer}, {Yoon}, \&
  {Heger}}]{Pet2005}
{Petrovic}, J., {Langer}, N., {Yoon}, S.-C., \& {Heger}, A. 2005, \aap, 435,
  247

\bibitem[{{Rashid} {et~al.}(2008){Rashid}, {Jones}, \& {Tobias}}]{Ras2008}
{Rashid}, F.~Q., {Jones}, C.~A., \& {Tobias}, S.~M. 2008, \aap, 488, 819

\bibitem[{{Rosenbluth} \& {Simon}(1964)}]{rs64}
{Rosenbluth}, M.~N., \& {Simon}, A. 1964, {Phys. Fluids}, 7, 557

\bibitem[{{Schatzman} {et~al.}(2000){Schatzman}, {Zahn}, \& {Morel}}]{Sch2000}
{Schatzman}, E., {Zahn}, J., \& {Morel}, P. 2000, \aap, 364, 876

\bibitem[{{Schmitt} \& {Rosner}(1983)}]{SR83}
{Schmitt}, Â.~H.~M.~M., \& {Rosner}, Â. 1983, \apj, 265, 901

\bibitem[{{Silvers} {et~al.}(2009){Silvers}, {Vasil}, {Brummell}, \&
  {Proctor}}]{Sil2009}
{Silvers}, L.~J., {Vasil}, G.~M., {Brummell}, N.~H., \& {Proctor}, M.~R.~E.
  2009, \apjl, 702, L14

\bibitem[{{Spruit}(1999)}]{Spr2000}
{Spruit}, H.~C. 1999, \aap, 349, 189

\bibitem[{{Talon} \& {Zahn}(1997)}]{TalZah1997}
{Talon}, S., \& {Zahn}, J.-P. 1997, \aap, 317, 749

\bibitem[{{Talon} {et~al.}(1997){Talon}, {Zahn}, {Maeder}, \&
  {Meynet}}]{Tal1997}
{Talon}, S., {Zahn}, J.-P., {Maeder}, A., \& {Meynet}, G. 1997, \aap, 322, 209

\bibitem[{{Tatsuno} \& {Dorland}(2006)}]{td06}
{Tatsuno}, T., \& {Dorland}, W. 2006, {Phys. Plasmas}, 13, 092107

\bibitem[{{Tatsuno} {et~al.}(2003){Tatsuno}, {Yoshida}, \& {Mahajan}}]{TYM03}
{Tatsuno}, T., {Yoshida}, Z., \& {Mahajan}, S.~M. 2003, {Phys. Plasmas}, 10,
  2278

\bibitem[{{Vasil} \& {Brummell}(2009)}]{VasBru2009}
{Vasil}, G.~M., \& {Brummell}, N.~H. 2009, \apj, 690, 783

\bibitem[{{Zahn}(1974)}]{Zah1974}
{Zahn}, J. 1974, in Stellar instability and evolution, ed. {P.~Ledoux,
  A.~Noels, \& A.~W.~Rodgers}, 185--194

\bibitem[{{Zahn}(1992)}]{Zah1992}
{Zahn}, J.-P. 1992, \aap, 265, 115

\end{thebibliography}

\end{document}